\documentclass[draftcls,onecolumn,12pt]{IEEEtran}%
\ifCLASSINFOpdf

\else
\fi

\usepackage{graphicx}
\usepackage{booktabs}
\usepackage{subcaption}
\usepackage{amsmath, array, makecell}
\allowdisplaybreaks[1]
\usepackage{amssymb}
\usepackage{eqnarray}
\usepackage{boldline, multirow}
\usepackage{amsbsy}
\usepackage{cite}
\usepackage{algpseudocode,algorithm,algorithmicx}
\usepackage[capitalise]{cleveref}
\usepackage[british]{babel}
\usepackage{csquotes}
\usepackage[dvipsnames]{xcolor}
\begin{document}
	\title{Hybrid Combining Based on Constant Phase Shifters and Active/Inactive Switches}
	\author{Eduard~E.~Bahingayi and
		Kyungchun~Lee,~\IEEEmembership{Senior Member,~IEEE}
		\thanks{E. E. Bahingayi and K. Lee are with the Department of Electrical and Information Engineering and the Research Center for Electrical and Information Technology, Nowon-gu, Seoul, 01811, Republic of Korea (e-mail: \{eduardbahingaye, kclee\}@seoultech.ac.kr).}
}
	
	\maketitle
	
	\begin{abstract}
		In this paper, we propose a new hybrid analog and digital combining architecture for millimeter wave (mmWave) multi-user multiple-input multiple-output (MU-MIMO) systems. The proposed structure employs antenna subset selection per radio frequency (RF) chain based on \textit{active}/\textit{inactive} switches and uses constant phase shifters (CPS) to control the phases of signals in the RF circuit. In this scheme, for each RF chain, a subset of receive antennas that contribute more to the desired signal power than the interference power is chosen for signal combining in the analog domain, whereas other receive antennas are excluded from signal combining, thereby enhancing sum-rates. Simultaneously, the proposed structure reduces power consumption in the RF circuit by exclusively activating switches that correspond to the antennas selected for each RF chain. We also develop three low-complexity algorithms for per-RF chain antenna subset selection. Finally, through numerical simulation, we show that the proposed structure provides higher spectral efficiency and higher energy efficiency than conventional hybrid analog and digital combining schemes for mmWave MU-MIMO systems. 
	\end{abstract}
	
	\IEEEpeerreviewmaketitle
	
	\section{Introduction}\label{S: I}
	
	\IEEEPARstart{D}{uring} the last decade, motivated by the potential of using the millimeter wave (mmWave) frequency for future mobile communication systems, mmWave massive multiple-input multiple-output (MIMO) have become a major research topic in the field of wireless communication\cite{Khan1,Rapp}. The benefits of mmWave massive MIMO systems are presented in \cite{Rapp,Sun,Ayanoglu}. However, its practical implementation remains problematic owing to the high hardware complexity and high power consumption of numerous RF chains, which scales with the number of antennas in conventional massive MIMO systems \cite{Larsson, Doan}. To address this problem, massive MIMO architecture for mmWave communication systems must be developed that uses fewer RF chains and low-power RF hardware components while still providing high data rates. \par	
	Antenna selection and hybrid beamforming schemes are often considered to be low-power solutions for reducing the number of RF chains in mmWave massive MIMO systems \cite{Zhang,Rusu,Sohrabi,Ayach,Rowell,Heath}. Between these two schemes, antenna selection is less energy-hungry because fewer antennas are selected and connected to the RF chains in antenna selection \cite{Rusu}. However, the loss of array gains and low data rates make this scheme less desirable. In contrast, the hybrid analog/digital (A/D) beamforming scheme has been deemed a better alternative owing to its use of fewer RF chains and its ability to provide high spectral efficiency (SE) close to that of the fully digital (FD) approach \cite{Sohrabi,Ayach,Rusu,Rowell,Abbas,Heath}. The hybrid A/D beamforming schemes can be classified into fully connected and sub-connected architectures. In a fully connected architecture, each RF chain is connected to all the antennas\cite{Ayach,Rusu}, whereas in a sub-connected architecture, each RF chain is connected to a subset of antennas \cite{Rusu,Zhu,Masouros}. The fully connected hybrid beamforming architecture outperforms the sub-connected architecture in terms of achievable throughput. \par	
	The main drawback of the fully connected hybrid beamforming architecture is its high power consumption due to a large number of variable phase shifters (VPS), which are used to implement the analog domain \cite{Rusu}. When the fully connected hybrid beamforming architecture is implemented in massive MIMO systems employing hundreds of antennas, the number of required VPSs can amount to more than a thousand \cite{Rowell,Ayach,Rusu,Sohrabi}, which can result in high power consumption \cite{Rusu,Graauw}. On combining all of these factors, the fully connected hybrid architecture can be less energy-efficient than FD schemes, particularly when the number of RF chains is greater than four \cite{Rusu,Abbas}. \par	
	The authors in \cite{Alkhateeb} proposed a novel fully connected hybrid A/D combining architecture in which the VPSs are replaced with constant phase shifters (CPS) and arrays of switches to design an analog beamformer. Using low-power CPSs to control signal phases in the analog domain, this architecture provides improved energy efficiency (EE) performance and a slight sum-rate loss compared to VPS-based hybrid A/D beamforming structures \cite{Alkhateeb,Buzzi}. Using the same structure, \cite{Sah} has presented novel algorithms based on quasi-orthogonal combining to maximize the signal-to-interference-plus-noise ratio (SINR) by reducing the interference power.\par	
	Both hybrid combining schemes presented in \cite{Alkhateeb} and \cite{Sah} for massive MIMO receivers achieve performance gains over the antenna selection scheme in terms of sum-rate. However, when the signals received at multiple receive antennas are combined at the RF chain, a certain subset of antennas can contribute more to the interference power than to the desired signal power depending on the channel conditions, which can cause loss in the SINR. In addition, owing to the large number of receive antennas in massive MIMO, the number of switches required for connecting antennas to the RF chains in the architecture proposed in \cite{Alkhateeb} is huge, and these switches can collectively consume a large amount of power.\par	
	To resolve the aforementioned problems, this study proposes a new hybrid analog and digital combining architecture for mmWave multi-user MIMO (MU-MIMO) receivers. The hardware structure of the proposed architecture is similar to that proposed in \cite{Alkhateeb}, except that the proposed architecture employs \textit{active}/\textit{inactive} switches in the RF circuit. Instead of connecting all switches to the CPSs, as in the scheme presented in \cite{Alkhateeb}, the proposed structure can set some of the switches to inactive states, thereby increasing the flexibility of the switching network. Furthermore, the use of active/inactive switches enables us to perform antenna subset selection for each RF chain. In particular, for each RF chain of the proposed architecture, only a subset of receive antennas are chosen---namely, those that contribute more to the desired signal power than to the interference power---and their signals are combined in the analog domain to enhance the achievable sum-rate while reducing power consumption. The recently published work in \cite{Payami} employs an array of switches to select a subset of VPSs, which are connected to the RF chains in the fully and sub-connected hybrid beamforming architectures. Unlike that in \cite{Payami}, in the proposed scheme, optimizing a switching network to design the RF combiners includes selecting the subset of antennas and their corresponding CPSs for each RF chain. Moreover, in the proposed structure, we reduce the number of active switches for each RF chain by modifying the fully connected CPS and switch (FCPS) architecture proposed in \cite{Alkhateeb}. \par 	
	To the best of our knowledge, previous studies on hybrid combining did not consider the possibility that a subset of antennas can cause a loss in the SINR owing to their larger contributions to interference power than desired signal power, which can degrade performance in terms of sum-rate. Traditional antenna subset selection schemes for massive MIMO systems are designed to select the \enquote*{best} subset of antennas that provides the optimal capacity close to what can be achieved when all antennas are used \cite{Rusu,Zhang,Gharavi}. In contrast, the proposed scheme selects the subset of antennas for each RF chain such that the proposed scheme outperforms conventional schemes without antenna selection.\par 	
	The proposed scheme has three advantages. First, the deactivation of switches enables reduced power consumption in a switch network because only the active switches are considered to consume power \cite{Rusu,Payami}. Second, the flexibility to exclude or include a subset of antennas in signal combining for each RF chain provides a higher degree of freedom for the design of RF combiners, thereby improving the SE of the proposed scheme compared to that of conventional schemes, where all antennas on each RF chain are in operation. Third, based on the first and second advantages, the proposed structure attains improved EE. The main contributions of this study can be summarized as follows:
	\begin{enumerate}
		\item We propose a new hybrid combining scheme for mmWave MIMO uplink systems that employs antenna subset selection per RF chain based on active/inactive switches and CPSs. In the proposed architecture, the per-RF chain antenna subset selection is achieved by activating only a subset of switches for each RF chain that corresponds to the selected antennas. This reduces the power consumption of the switching network compared to that of the architecture presented in \cite{Alkhateeb}. Moreover, the use of active/inactive switches in the proposed architecture offers a high degree of freedom to design RF combiners, thereby enhancing the achievable SE.  
		\item We develop three near-optimal algorithms for the antenna subset selection and hybrid combining of the proposed scheme. First, we investigate a system employing an arbitrary number of active switches for each RF chain, where the subset of antennas for each RF chain is selected through a decremental search-based algorithm to maximize the achievable SE of the system. Then, to reduce the complexity, we develop two algorithms that employ the same number of active switches for each RF chain, where the subset of antennas for each RF chain is selected based on the channel magnitude.
		\item For the proposed structure, extensive performance comparisons based on numerical results are provided to reveal that the proposed per-RF chain antenna subset selection scheme can attain higher SE and EE as compared with conventional hybrid combining schemes.
	\end{enumerate}
	The remaining paper is organized as follows. In \cref{S: II}, we provide the system model of the proposed MU-MIMO receiver and the channel model used for our study. In \cref{S: III}, we present the algorithms used to perform the per-RF chain antenna subset selection. \cref{S: IV} provides information on the EE of the proposed scheme in comparison with conventional structures. In \cref{S: V}, simulation results are presented to numerically evaluate the proposed scheme. Finally, \cref{S: VI} presents some concluding remarks.\par
	The following notations are used throughout this article: a boldface capital letter, $ {\textbf{X}}$, is used to denote a matrix, and a boldface lowercase letter, ${\textbf{x}}$, denotes a column vector. The $ {n} $th entry of vector $ {\textbf{x}} $ is denoted by $ [\textbf{x}]_{n}$ or $x_{n}$. The $ {n} $th row and $ {m} $th column entry of $ \textbf{X}$ is denoted by $[\textbf{X}]_{n,m} $ or $ x_{n,m} $. We also use $\textbf{X}^{\text{H}}$, $\textbf{X}^{\text{T}}$, and $\textbf{X}^{-1}$ to denote the hermitian transpose, transpose, and inverse of $ {\textbf{X}}$, respectively. $\mathrm{diag}\ \big[\textbf{X}_1,\textbf{X}_2,\cdots,\textbf{X}_i \big]$ is a block diagonal matrix containing $ \textbf{X}_1,\cdots,\textbf{X}_i $ as its diagonal terms. $\left\Vert \textbf{X} \right\Vert_{F} $ is the Frobenius norm of ${\textbf{X}}$; $ \left\Vert \textbf{x} \right\Vert_{0} $ is the ${l}_{0}$ pseudo-norm of $ \textbf{x}$; $\left\vert{x}\right\vert$ is the magnitude of scalar $ {x} $; and $\textbf{X} \odot \textbf{Y} $ is the element-wise multiplication of $ \textbf{X} $ and $ \textbf{Y} $. The calligraphic letter, $\mathcal{X}$, denotes a set, and $\left\vert\mathcal{X}\right\vert$ represents the cardinality of set $\mathcal{X}$. Finally, we use $ \textbf{1}_{N \times M} $ and $ \textbf{0}_{N \times M} $ to denote the ${N \times M} $ matrix with all one and zero entries, respectively.	
	\section{System Model}\label{S: II}	
	\begin{figure}[!ht]
		\centering
		\includegraphics[scale = 0.63]{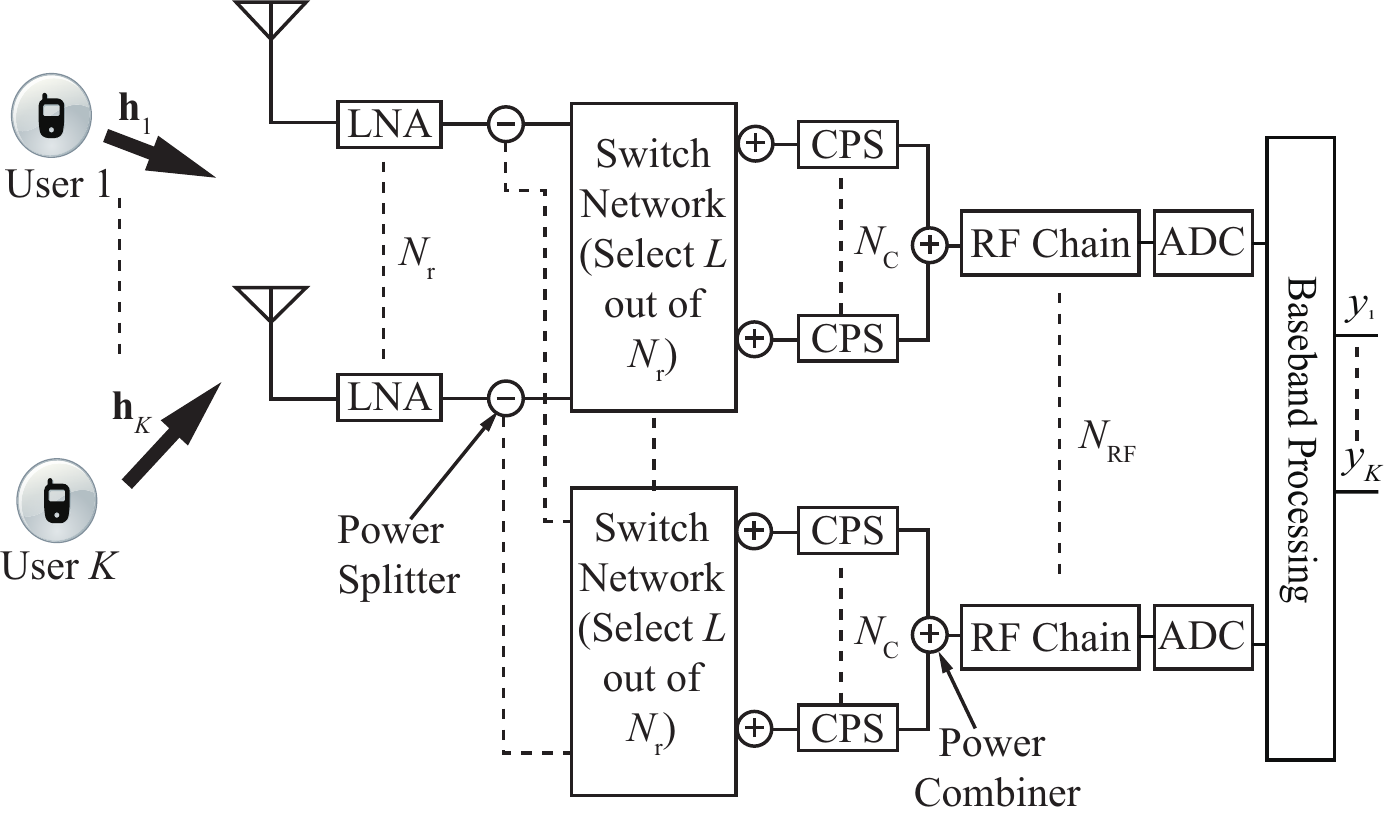}
		\caption{System model.} 
		\label{fig:1}
	\end{figure}
	\begin{figure}[!ht]
		\centering
		\includegraphics[scale = 0.43]{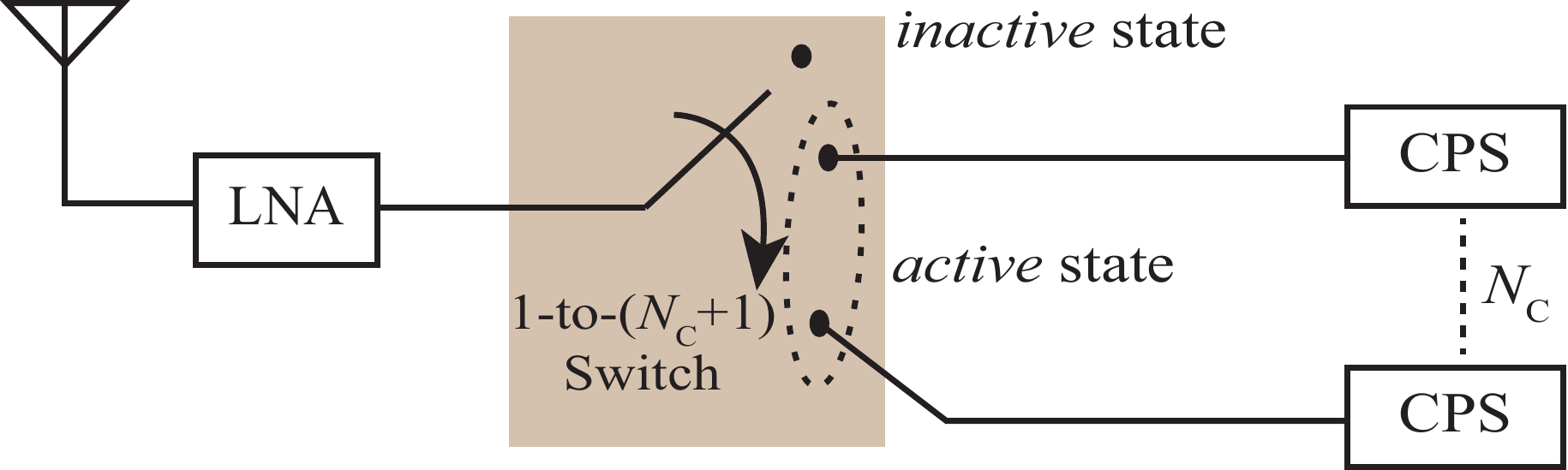}
		\caption{ An antenna-to-CPS switch in a switch network.}
		\label{fig:2} 
	\end{figure}
	\raggedbottom	
	\subsection{Hybrid Combining Receiver} \label{section:IIA}	
	\cref{fig:1} illustrates the system model including the proposed MU-MIMO receiver structure. We assume that $ {K} $ single-antenna users transmit their signals to the base station (BS) \cite{Choi,Alkhateeb}, which has ${N}_{\text{r}}$ receive antennas and ${N}_{\text{RF}} (\geq {K})$ RF chains. For simplicity, we assume that the BS employs exactly $K$ out of ${N}_{\text{RF}}$ available RF chains to simultaneously serve $K$ users, as was assumed in \cite{Alkhateeb,Heath,Wang}. The power splitter is used after the receive antennas to distribute the received signal to multiple RF chains. We assume that each RF chain is connected to ${L}$ $ (\leq {N}_{\text{r}})$ antennas, which are selected by a switch network through ${N}_{\text{C}}$ CPSs. The signals of the selected receive antennas are added by power combiners. Considering an antenna-to-CPS connection, each switch can be active or inactive, as shown in \cref{fig:2}. Consequently, in a switch network, an antenna is connected to a CPS only when its switch is active. Conversely, if a switch is inactive, it implies that the corresponding antenna is not connected to any CPS.\par
	The received signal at the front end of the BS is given as follows:
	\begin{align}
	\textbf{r} = \sqrt{p} {\textbf{H}}{\textbf{x}}+ {\textbf{n}},
	\end{align}
	where ${\textbf{r }\in \mathbb{C}^{N_{\text{r}} \times {1}}} $ is the received signal vector; $p$ is the average received power from all users; and $ \textbf{x}=\left[x_1,x_2,\cdots,x_K\right]\in \mathbb{C}^{K \times {1}} $ is a transmitted signal vector, where $ {x_{k}}$ is the symbol transmitted by the $ {k} $th user. Furthermore, ${\textbf{H}=\left[\textbf {h}_1,\textbf{h}_2,\cdots,\textbf{h}_K\right] \in \mathbb{C}^{N_{\text{r}} \times {K}}} $ denotes the channel matrix, where ${\textbf {h}_k}\in \mathbb{C}^{N_{\text{r}} \times {1}}$ is the channel vector of the user $ {k} $, whereas ${\textbf{n} \in \mathbb{C}^{N_{\text{r}} \times {1}}}$ denotes the independent and identically distributed (i.i.d) additive white Gaussian noise vector with $n_{i}\sim\mathcal{CN}(0,\sigma^2)$. The received signal can be rewritten as
	\begin{align}\label{recsig} 
	{\textbf{r}} = \sqrt{p}{\textbf{h}_k}{{x}_k}+\sqrt{p}\sum\limits_{i\neq k} \textbf{h}_i{{x}_i}+ {\textbf{n}}. 
	\end{align}
	To detect $ {{x}_k} $, the receiver applies hybrid analog and digital combining to the received signal, which generates 
	\begin{align}
	{{y}}_k =\sqrt{p}{ \textbf{w}_{\text{HBF},{k}}^{\text{H}}}{ \textbf{h}_k}{\textbf{x}_k}+\sqrt{p}{ \textbf{w}_{\text{HBF},{k}}^{\text{H}}}\sum\limits_{i\neq k} \textbf{h}_i{\textbf{x}_i}+ { \textbf{w}_{\text{HBF},{k}}^{\text{H}}}{\textbf{n}}, 
	\end{align}
	where ${ \textbf{w}_{\text{HBF},{k}}}={\textbf{W}_\text{RF}} \textbf{w}_{\text{BB},{k}} \in \mathbb{C}^ {{N_{\text{r}}} \times {1}}$ is the hybrid combining vector for $ {{x}_k} $, whereas ${\textbf{W}_\text{RF}}\in \mathbb{C}^{N_{\text{r}} \times {K}}$ is the analog beamforming (ABF) matrix, and $\textbf{w}_{\text{BB},{k}}\in \mathbb{C}^{K\times {1}}$ is the digital combining vector for $ {{x}_k} $.\par
	In the proposed structure, the ABF matrix is implemented using switches and CPSs, where CPSs are subject to constant modulus. The ABF matrix can be expressed as 
	\begin{align}\label{ABF}
	{\textbf{W}_\text{RF}} = \left[{\boldsymbol\Delta_{1}} {\boldsymbol\psi}, {\boldsymbol\Delta_{2}} {\boldsymbol\psi},\cdots, {\boldsymbol\Delta_{K}} {\boldsymbol\psi}\right],
	\end{align}
	where $ \boldsymbol{\psi} =[1,e^{j\frac{2\pi}{N_{\text{C}}}},\cdots,e^{j\frac{{2\pi}{(N_{\text{C}} -1)}}{N_{\text{C}}}}]^{\text{T}} $ represents an array of $ N_{\text{C}} $ possible constant phases. The composite switching matrix is represented by ${\boldsymbol\Delta}=\left[{\boldsymbol\Delta_{1}}, {\boldsymbol\Delta_{2}},\cdots,{\boldsymbol\Delta_{K}}\right]\in \mathcal{B}^{{N_{\text{r}}} \times {N_{\text{C}}K}}$, where ${\boldsymbol\Delta_{k}}\in\mathcal{B}^{{N_{\text{r}}} \times {N_{\text{C}}}}$ is the switching matrix for the $ {k} $th RF chain that satisfies the following constraints:
	\begin{subequations} \label{cns}
		\begin{align}
		\begin{split}\label{cns:1}
		[{\boldsymbol\Delta_{k}}]_{n,q}&\in {\{0,1}\},\hspace{1mm}\forall {n,q},
		\end{split}\\
		\begin{split}\label{cns:2}
		\sum\limits_{q=1}^{N_{\text{C}}}[{\boldsymbol\Delta_{k}}]_{n,q}&\in{\{0,1}\},
		\end{split}
		\end{align}
	\end{subequations}
where $ {n=1,\cdots,{N_{\text{r}}}} $ represents the antenna index and $ {q=1,\cdots,{N_{\text{C}}}}$ is the CPS index. The constraint in (\ref{cns:1}) represents the use of switches, and (\ref{cns:2}) implies that the restriction of each antenna on each RF chain is connected to at most one CPS. For instance, in a system with $ N_{\text{r}} = 6 $ and $ N_{\text{C}} = 3 $, an example of the switching matrix $ {\boldsymbol\Delta_{k}}$ can be
	\begin{equation} \label{meq}
	{\boldsymbol\Delta_{k}}=\begin{bmatrix}
	1 & 0 & 0 \\
	0 & 0 & 0 \\
	0 & 1 & 0 \\
	0 & 0 & 1 \\
	0 & 0 & 0 \\
	1 & 0 & 0 \\
	\end{bmatrix}.
	\end{equation}	
	In (\ref{meq}), the second and fifth rows indicate that the switches for antennas $ 2 $ and $ 5 $ are inactive for the $ {k} $th RF chain, whereas each of the other antennas is connected to a CPS. Thus, the signals of antennas $ 2 $ and $ 5 $ are excluded from signal combining for the $ {k} $th RF chain, whereas other signals are selected for combining. When $ {L}_{k} $ represents the number of active switches for the $ {k} $th RF chain, we have 
	\begin{equation} \label{cns:3}
	\sum\limits_{q=1}^{N_{\text{C}}}\sum\limits_{n=1}^{N_{\text{r}}}[{\boldsymbol\Delta_{k}}]_{n,q}= L_k ,\hspace{1mm}1\leq L_k \leq N_{\text{r}}.
	\end{equation}
	In the example presented in (\ref{meq}), $ L_{k} $ is equal to four. The $ {k} $th column vector of an ABF matrix is given by ${\textbf{w}_{\text{RF},k}}={\boldsymbol\Delta_{k}} {\boldsymbol{\psi}}$. It has non-zero entries corresponding to the antennas connected to the active state switches. In contrast, if an antenna is connected to an inactive switch, its corresponding entry in ${\textbf{w}_{\text{RF},k}}$ becomes zero.\par 
	Therefore, in the proposed architecture, each entry of $ {\textbf{W}_\text{RF}} $ can be either 0 or of unit modulus, i.e., $\vert{{w}_{i,j}\vert \in {\{0,1}\} } $. The number of selected antennas for the $ {k} $th RF chain is given by $ {\|{\textbf{w}}_{\text{RF},k}}\|_0 = L_{k}$, where we have $ 1 \leq L_{k} \leq N_{\text{r}} $, $ k=1,2,\cdots,K$.
	The SINR of the $ {k} $th user at the BS can be expressed as
	\begin{align}\label{SNR}
	\text{SINR}_k=\frac{{p}\left\vert {\textbf{w}_{\text{BB},k}^{\text{H}}}{\textbf{W}_{\text{RF}}^{\text{H}}}{\textbf{h}_k}\right\vert ^2} {{p}\sum_{i \neq k}^{K} \left\vert {\textbf{w}_{\text{BB},k}^{\text{H}}}{\textbf{W}_{\text{RF}}^{\text{H}}}{\textbf{h}_i}\right\vert ^2+\sigma^2 \left\Vert {\textbf{w}_{\text{BB},k}^{\text{H}}}{\textbf{W}_{\text{RF}}^{\text{H}}}\right\Vert^2}.
	\end{align}
	Our aim is to design the RF and digital combiners ${\{\textbf{W}_{\text{BB}}}, {\textbf{W}_{\text{RF}}}\} $ in such a manner as to maximize the overall achievable sum-rate of the uplink MU-MIMO system in \cref{fig:1}, which can be formulated as
	\begin{subequations} \label{rate:1}
		\begin{align}
		\begin{split}
		{C} = \displaystyle \max_{{\textbf{W}_{\text{RF}}}\in \mathcal{W},\textbf{W}_{\text{BB}}}&{\sum\limits_{k=1}^{K} {\log (1+ \text{SINR}_k)}},		
		\end{split}\\
		\begin{split}
		\textrm{subject to} \hspace{1mm} &\left\vert{[\textbf{W}_{\text{RF}}}]_{i,j}\right\vert \in {\{0,1}\},
		\end{split}
		\end{align}
	\end{subequations} 
	where $\mathcal{W}$ represents a set of ABF matrices satisfying (\ref{ABF}), (\ref{cns:1}), and (\ref{cns:2}).
	\subsection{Channel Model} 
	In this study, we employ a geometric channel model as a propagation environment between each user terminal and the BS, which is a typical channel model assumed for mmWave massive MIMO systems \cite{Liang,Sohrabi,Choi}. We assume that the channel of each user has an equal number of independent propagation paths $ {N_{\text{p}}}$ \cite{Sohrabi,Liang}. The channel vector between the $ {k} $th user and the BS is given by
	\begin{equation}
	{\textbf{h}_k} = \sqrt{\frac{N_{\text{r}}}{N_{\text{p}}}} \sum\limits_{l=1}^{N_{\text{p}}} \textbf{a}({\phi_l^k}){\alpha_l^k},
	\end{equation}
	where $\alpha_l^k \sim\mathcal{CN}(0,1)$ is the complex gain of the $ l $th path; ${\phi_l^k}\in[0,2\pi]$ denotes the angle of arrival (AoA) of the $ l $th path; and $ \textbf{a}(\cdot) $ represents the antenna array response vector at the BS. We also assume that the BS is equipped with a uniform linear array, for which the array response vector can be modeled as \cite{Sohrabi,Xiao,Heath}
	\begin{align}{\textbf{a}(\phi)} = \frac{1}{\sqrt{N_{\text{r}}}}\left[1,e^{j2\pi\frac{d}{\lambda}\sin (\phi)},\cdots ,e^{j2\pi (N_{\text{r}}-1)\frac{d}{\lambda}\sin (\phi) }\right]^{\text{T}},
	\end{align}
	where ${d}$ is the antenna spacing, and $\lambda$ is the wavelength of the carrier signal.
	
	\section{Antenna Subset Selection and Hybrid Combining Design}\label{S: III}
	As described in the previous section, in the proposed architecture, RF combining is implemented using switches and CPSs. Thus, to obtain the optimal RF combiner in (\ref{rate:1}), we must solve the combinatorial problem of designing a switching matrix, which poses two subproblems. The first problem is how to determine, in a search across $ 2^{N_{\text{r}}{K}}$ possible combinations, the states of $ {N_{\text{r}}{K}} $ switches, which can be combinatorially prohibitive when $ {N_{\text{r}}} $ is very large. The second combinatorial problem is determining the optimal connection between the selected antennas and $ {N_{\text{C}}} $ available CPSs, given that, if $ {L} $ out of $ {N_{\text{r}}} $ antennas are selected for each RF chain, then there are ${N_{\text{C}}}^{{K}{L}}$ possible connections between the selected antennas and $ {N_{\text{C}}} $ available CPSs. Therefore, herein, we solve the aforementioned subproblems by designing the switching matrix in two stages based on low complexity algorithms, which can provide near-optimal solutions. \par
	In the first stage, we ignore the antenna subset selection. The switching matrices in the first stage corresponding to $ \boldsymbol\Delta $ and $ \boldsymbol\Delta_{k} $ are denoted by $\tilde{\boldsymbol\Delta}\in \mathcal{B}^{{N_{\text{r}}} \times {N_{\text{C}}K}}$ and ${\tilde{\boldsymbol\Delta}_{k}}\in\mathcal{B}^{{N_{\text{r}}} \times {N_{\text{C}}}}$, respectively. The constraints (\ref{cns:2}) and (\ref{cns:3}) are modified to 
	\begin{subequations} \label{cns_mod}
		\begin{align}
		\begin{split}\label{cns_mod:2}
		\sum\limits_{q=1}^{N_C}[{\tilde{\boldsymbol\Delta}_{k}}]_{n,q}&={1}
		\end{split},\\
		\begin{split}\label{cns_mod:3}
		\sum\limits_{q=1}^{N_{\text{C}}}\sum\limits_{n=1}^{N_{\text{r}}}[{\tilde{\boldsymbol\Delta}_{k}}]_{n,q}&= N_{\text{r}}
		\end{split}.
		\end{align}
	\end{subequations}
	Then, the ABF matrix $\textbf{W}_{\text{RF}}$ has no zero entries because all switches are active. For simplicity of notation, this ABF matrix with no zero entries is denoted by $ \tilde{\textbf{W}}_{\text{RF}}\in \mathbb{C}^{N_{\text{r}} \times {K}} $. \par 	
	Seeking a low-complexity solution, we adopt the Euclidean distance method in \cite{Alkhateeb,Liang} to design a switching matrix $\tilde{\boldsymbol\Delta}$. Furthermore, by using QR decomposition \cite{Golub}, we can express ${\textbf{H}} = \hat{\textbf{H}}\textbf{R}$, where $\hat{\textbf{H}} $ of size $ N_{\text{r}} \times K $ forms an orthonormal set of basis vectors for the column space of $\textbf{H}$, and $ \textbf{R}$ of size $K \times K $ is an upper-triangular matrix. In this study, we exploit $\hat{\textbf{H}} $ to generate $ \tilde{\textbf{W}}_{\text{RF}}$. The advantage of using $\hat{\textbf{H}} $ over ${\textbf{H}} $ is that the column vectors of $ \tilde{\textbf{W}}_{\text{RF}}$ become approximately collinear to the corresponding column vectors in $ {\textbf{H}}$. As a consequence, in (8), the inner product of approximately collinear vectors in the numerator term generated by the desired signal can be enhanced, whereas the inner product of near-orthogonal vectors in the first term of the denominator generated by interference signals is reduced, thereby enhancing the SINR \cite{Sah,Xiao}. In this scheme, based on the shortest Euclidean distance, the $ {n} $th antenna's switch in $ {\tilde{\boldsymbol\Delta}_{k}} $ selects the CPS with phase $ \hat{\theta}_{k,n} $ from $ \boldsymbol\psi $, which corresponds to the closest phase of $ [\hat {\textbf{h}}_{k}]_{n} $ \cite{Liang}. Letting $\theta_{k,n}=\angle [\hat {\textbf{h}}_{k}]_{n} $ denote the phase of the channel coefficient corresponding to the $ {n} $th antenna on the $ {k} $th RF chain, we obtain 
	\begin{equation}\label{stage1:1}
	\hat{\theta}_{k,n} = \frac{{2\pi(\hat{q}-1)}}{N_{\text{C}}}, 
	\end{equation}
	where $\hat{q}= \arg\underset{q\in{\{ 1,2,\cdots,N_{\text{C}}\}}}\min \left\vert \theta_{k,n} - \frac{2\pi (q-1)}{N_{\text{C}}}\right\vert $ is the index of the chosen CPS. Then, the corresponding switch is set to the active state, i.e.,
	\begin{equation}\label{stage1:2}
	[{\tilde{\boldsymbol\Delta}_{k}}]_{n,\hat{q}}=1.
	\end{equation}
	Consequently, the $ {k} $th column vector of $ \tilde{\textbf{W}}_{\text{RF}}$ becomes
	\begin{equation}\label{stage1:3}
	{\tilde{\textbf{w}}_{\text{RF},k}} = {\tilde{\boldsymbol\Delta}_{k}} \boldsymbol\psi, \hspace{1mm} k=1,2,\cdots,K.
	\end{equation} \par
	In (\ref{stage1:1}) and (\ref{stage1:2}), the switching matrix is designed under the assumption that no antenna subset selection is performed, implying that all switches are active. However, the main objective in (\ref{rate:1}) is to design $ \textbf{W}_{\text{RF}} $ under the constraints in (\ref{cns}), which implies that only a set of selected switches are put into active states. Therefore, by considering the constraints in (\ref{cns}), we must modify the switching matrix obtained in the first stage, which leads us to the following stage.\par 
	In the second stage, we introduce a matrix $ \textbf{S} = [\textbf{s}_{1},\cdots,\textbf{s}_{K}] \in {\mathcal{B}}^{N_{\text{r}}\times K}$. In each column vector, $ \textbf{s}_{k} $, $ 1 \mathrm{s} $ correspond to the antennas selected for the $ {k} $th RF chain. We note that the subset of antennas selected for the $ {k} $th RF chain is not necessarily the same as that selected for the $ {j} $th RF chain, $ {j} \neq {k} $. For example, $ \mathcal{S}_{1} ={\{1,3,4,6}\}$ can be a set of antenna indices selected for the first RF chain, whereas, simultaneously, $ \mathcal{S}_{2} ={\{1,2,4,5,6}\}$ are selected for the second RF chain. Hence, when the $ {n} $th antenna is not selected for the $ {k} $th RF chain, it does not mean the $ {n} $th antenna is totally inactive because it can be selected for some other RF chain.\par	
	Here, $ \textbf{S} $ transforms $ \tilde{\boldsymbol\Delta} $ into $ \boldsymbol\Delta $ such that the antenna subset selection is considered. The $ {k} $th column of $\textbf{S}$ is element-wise multiplied by each column of $\tilde{\boldsymbol\Delta}_{k}$ to generate ${\boldsymbol\Delta}_{k}$. Specifically, for ${\tilde{\boldsymbol\Delta}_{k}} =[\tilde{\boldsymbol\delta}_{1},\tilde{\boldsymbol\delta}_{2},\cdots,\tilde{\boldsymbol\delta}_{N_{\text{C}}}]$, we obtain
	\begin{equation} \label{sm:17}
	{\boldsymbol\Delta} = [\textbf{s}_{1} \odot \tilde{\boldsymbol\delta}_{1},\cdots,\textbf{s}_{1} \odot \tilde{\boldsymbol\delta}_{N_{\text{C}}}, \cdots,\textbf{s}_{K} \odot \tilde{\boldsymbol\delta}_{1},\cdots,\textbf{s}_{K} \odot \tilde{\boldsymbol\delta}_{N_{\text{C}}}].
	\end{equation}
	Based on (\ref{ABF}) and (\ref{sm:17}), the $ {k} $th column vector of the ABF matrix is expressed as
	\begin{equation}
	{\textbf{w}_{\text{RF},k}} = [\textbf{s}_{k} \odot \tilde{\boldsymbol\delta}_{1},\cdots,\textbf{s}_{k} \odot \tilde{\boldsymbol\delta}_{N_{\text{C}}}] \boldsymbol\psi,\\
	\end{equation}
	which can be simplified into
	\begin{equation} \label{eqABF}
	{\textbf{w}_{\text{RF},k}} = \textbf{s}_{k} \odot (\tilde{\boldsymbol\Delta}_{k} \boldsymbol\psi)=\textbf{s}_{k} \odot {\tilde{\textbf{w}}_{\text{RF},k}},
	\end{equation}
	thus yielding
	\begin{equation} \label{WRF:1}
	\textbf{W}_{\text{RF}} = [\textbf{s}_{1} \odot {\tilde{\textbf{w}}_{\text{RF},1}},\cdots,\textbf{s}_{K} \odot {\tilde{\textbf{w}}_{\text{RF},K}}] =\textbf{ S} \odot {\tilde{\textbf{W}}_{\text{RF}}}.
	\end{equation}
	Consequently, (\ref{WRF:1}) suggests that $ \textbf{W}_{\text{RF}} $ can be obtained via element-wise multiplication of $ {\tilde{\textbf{W}}_{\text{RF}}}$ and $ \textbf{S} $, where ${\tilde{\textbf{W}}_{\text{RF}}} $ is created by quantizing each phase entry of $\hat{\textbf{H}}$ to the nearest phase among the $ N_{\text{C}} $ possible phases of CPSs, while $ \textbf{S} $ contains $ 1 \mathrm{s} $ in the entries corresponding to the selected antennas for each RF chain and $0\mathrm{s}$ elsewhere.\par 
	For baseband digital combining, we adopt the minimum mean squared error (MMSE) beamforming scheme, which utilizes the effective channel $ \textbf{H}_{\textnormal{e}} = \textbf{W}_{\text{RF}}^{\text{H}} {\textbf{H}} $ \cite{Bengtsson,Wang}. Specifically, the  digital combiner of the $ {k} $th user can be written as
	\begin{equation} \label{WBB1}
	{{\textbf{w}}_{\text{BB},k}}={({\textbf{I}}_K + {\dfrac{{p}}{\sigma^2}}({\textbf{H}_{\textnormal{e}}^{\text{H}}}{\textbf{H}_{{\textnormal{e}}}}))^{-1} {\textbf{h}_{{{\textnormal{e}},k}}^{\text{H}}}}, 
	\end{equation}
	where ${\textbf{h}_{{{\textnormal{e}},k}}}\in \mathbb{C}^ {{N_{\text{r}}} \times {1}} $ is the $ {k} $th column vector of $ \textbf{H}_{\textnormal{e}} $.
	Using $ \textbf{W}_{\text{RF}} $ in (\ref {WRF:1}) and $ {\textbf{w}_{\text{BB},k}} $ in (\ref{WBB1}), the sum-rate of $ {K} $ users can be calculated as 
	\begin{equation} \label{Capacity:1}
	{C}={\sum\limits_{k=1}^{K} {\log (1+ \text{SINR}_k)}},
	\end{equation}
	where $ \text{SINR}_k $ represents the post-combining SINR of the $ {k} $th user, which is given by (\ref{SNR}). Thus, the optimization of $ \textbf{S} $ can be formulated as 
	\begin{subequations}\label{Capacity:2}
		\begin{align}
		\begin{split}
		{{\textbf{S}}^{\star}} =\arg\underset{\textbf{S}}\max &{\sum\limits_{k=1}^{K} {\log (1+\text{SINR}_k)}}		
		\end{split}\\
		\begin{split}
		\textrm{subject to} \hspace{1mm} &[\textbf{S}]_{n,k} \in {\{ 0,1 }\},\hspace{1mm}\forall_{n,k},
		\end{split}\\
		\begin{split}
		\left\Vert \textbf{s}_{k} \right\Vert_{0}& = L_{k};\hspace{1mm}1\leq L_{k} \leq N_{\text{r}}.\\
		\end{split}
		\end{align}
	\end{subequations}	
	As discussed earlier in this section, the antenna subset selection in (\ref{Capacity:2}) is a combinatorial problem, where the exhaustive search to find optimal $ \textbf{S} $ might entail excessive complexity. In the following subsections, we propose three near-optimal algorithms to perform antenna subset selection with significantly lower complexities.	 	
	\subsection{Decremental search-based ASS}
	In this subsection, we describe the proposed decremental search-based antenna subset selection (DS-ASS) scheme. In this scheme, we first assume that all antennas are connected to each RF chain and compute the initial sum-rate. Then, in every iteration, we search for an antenna on each RF chain that causes maximum increment in the sum-rate when it is disconnected. The chosen antenna in each iteration is removed from the set of antennas for that particular RF chain. Next, the search process is repeated until the maximum sum-rate is achieved. The sum-rate can decrease in some iterations and increase again before it reaches the maximum; hence, the search process continues up to $ {t_{max}}$ consecutive iterations, even if the sum-rate decreases at a certain step. Early termination occurs if the sum-rate decreases for more than $ {t_{max}}$ consecutive iterations.\par	
	\begin{algorithm}[!h]
		\caption{Proposed DS-ASS}
		\begin{algorithmic}[1]
			\renewcommand{\algorithmicrequire}{\textbf{Input:}}
			\renewcommand{\algorithmicensure}{\textbf{Output:}}
			\Require $ N_{\text{C}}$, $ \dfrac{p}{\sigma^2} $, $\textbf{H} $, and $\hat{\textbf{H}}$.
			\Ensure $ {{\textbf{S}}^{\star}} $.
			\\ \textit{Initialization} : $\textbf{\textbf{S}}=\textbf{1}_{N_{\text{r}} \times K}$
			\State Quantize the phase $ \theta $ of each entry of $ \hat{\textbf{H}} $ to the nearest possible phase $ \hat{\theta} $ and generate ${\tilde{\textbf{W}}_{\text{RF}}}$ based on (\ref{stage1:1})$ \-- $(\ref{stage1:3}).
			\State Compute the initial sum-rate $ c_{0} $ based on (\ref{WRF:1})$ \-- $(\ref{Capacity:1}),
			\State $ c_{best} \gets c_{0} $
			\State Reshape $ \textbf{S } $ into a vector $ \textbf{s}=[s_{1},s_{2},s_{3},\cdots,s_{N_{\text{r}}K}]^{\text{T}}$.
			\State $ \textbf{s}_{best} \gets \textbf{s} $
			\State $ t = 0 $
			\While {$ t < t_{max} $}
			\State $\mathcal{T} \ \gets $ indices of non-zero elements in $\textbf{s}$
			\State $ LocIter =\vert\mathcal{T}\vert$
			\For {$i = 1 $ to $LocIter$}
			\State ${\hat{\textbf{s}}}\gets \textbf{s}$
			\State $j\gets {i}$th element of $ \mathcal{T} $
			\State ${\hat{\textbf{s}}} _{j} = 0$
			\State Reshape $ {\hat{\textbf{s}}} $ to generate $\textbf{S}$.
			\State Compute the sum-rate $ c_{j} $ based on (\ref{WRF:1})$ \-- $(\ref{Capacity:1}). 
			\EndFor
			\State ${ j^{\star}}= \arg \underset{j\in\mathcal{T}}\max {\hspace{1 mm} c_{j}} $
			\State Update the index vector: $ [\textbf{s}]_{j^{\star}} = 0 $ 
			\If {$c_{j^{\star}} \geq c_{best} $}
			\State Update $ c_{best} \gets c_{j^{\star}} $, $ \textbf{s}_{best} \gets s $.
			\State $ t \gets 0 $ 
			\Else 
			\State $ t \gets t + 1 $.
			\EndIf
			\EndWhile
			\State Reshape $ \textbf{s}_{best} $ to generate $ {{\textbf{S}}^{\star}} $. \\			
			\Return $ {{\textbf{S}}^{\star}} $ 
		\end{algorithmic} 
		\label{alg:1}
	\end{algorithm}	
	The operations of DS-ASS are summarized in \cref{alg:1}. In step 1, $\textbf{S}$ is initialized as $ \textbf{1}_{N_{\text{r}} \times K} $. In step 3, we compute the initial sum-rate using (\ref{WRF:1})$ \-- $(\ref{Capacity:1}), which is set to the best sum-rate $ c_{best} $. Then, in step 5, $ \textbf{S} $ is reshaped into $ \textbf{s}= [s_{1},s_{2},s_{3}, \cdots,s_{N_{\text{r}}K}]^{\text{T}}$ by concatenating all columns of $ \textbf{S} $ into one column, which is also set to $ \textbf{s}_{best} $. As shown in steps $12\--14$, in each local iteration, a non-zero entry of $ \textbf{s}$ is converted to 0, which generates a new vector $ \hat{\textbf{s}} $. Then, $ \hat{\textbf{s}} $ is reshaped to $ \textbf{S} $, and in step 16, we compute the corresponding sum-rate $ c_{j} $. When the local iterations are finished, we obtain the index of the maximum sum-rate, $ j^{\star} $, in step 18. Then, in step 19, $ \textbf{s} $ is updated by replacing the $ j^{\star} $th entry with 0. Next, the maximum sum-rate $ c_{j^{\star}}$ is compared to the current largest sum-rate $ c_{best}$. If $ c_{j^{\star}} \geq c_{best}$, $ c_{best} $ and $ \textbf{s}_{best} $ are updated to $ c_{j^{\star}} $ and $\textbf{s}$, respectively, as shown in steps $ 20\-- 21 $, and the search process is repeated. If $ c_{j^{\star}} < c_{best}$ the search process can continue without updating $ c_{best} $ and $ \textbf{s}_{best} $ as long as $ t < t_{max} $. The search process is terminated early only if the sum-rate fails to increase for $ t_{max} $ consecutive iterations. Finally, in step 27, $\textbf{s}_{best} $ is reshaped to a matrix to generate $ \textbf{S}^{\star} $, which is the solution for antenna subset selection.
	
	\subsection{Channel magnitude-based ASS}
	In this subsection, we propose a low-complexity scheme, called channel magnitude-based antenna subset selection (CM-ASS), to further reduce the complexity of antenna subset selection. In the CM-ASS scheme, unlike DS-ASS, each RF chain is connected to the same number of active switches, i.e., $ {L_k}=L,\forall{_k} $. This scheme can be further divided into two different schemes. Specifically, we consider the cases when $ L $ is dynamic and when $ L $ is fixed. In the former case, $ L $ is obtained through an iterative process to maximize the sum-rate, and the value of $ L $ varies according to channel conditions. In contrast, in the latter case, $ L $ is fixed to a predefined value.\par
	\subsubsection{CM-ASS with dynamic L}
	In this scheme, we compute the initial sum-rate under the assumption that all antennas are connected to the RF chain. Then, for every iteration, $ K $ connections between antennas and RF chains are removed, one from the subset of antennas for each RF chain that corresponds to the entry with the smallest magnitude in each column of $ \hat{\textbf{H}} $. Next, the corresponding sum-rate is calculated. The search process is repeated if the new sum-rate is greater than the current largest sum-rate. Similar to the DS-ASS scheme, the CM-ASS scheme with dynamic $ L $ terminates the search process after $ t_{max} $ consecutive failures to increase the sum-rate.\par
	\begin{algorithm} [!t]
		\caption{Proposed CM-ASS with dynamic\textit{ L}}
		\begin{algorithmic}[1]
			\renewcommand{\algorithmicrequire}{\textbf{Input:}}
			\renewcommand{\algorithmicensure}{\textbf{Output:}}
			\Require $ N_{\text{C}}$, $ \dfrac{p}{\sigma^2} $, $\textbf{H} $, and $\hat{\textbf{H}}$.
			\Ensure $ {{\textbf{S}}^{\star}} $.
			\\ \textit{Initialization} : $\textbf{\textbf{S}}=\textbf{1}_{N_{\text{r}} \times K}$
			\State Quantize the phase $ \theta $ of each entry of $ \hat{\textbf{H}} $ to the nearest possible phase $ \hat{\theta} $ and generate ${\hat{\textbf{W}}_{\text{RF}}}$ based on (\ref{stage1:1})$ \-- $(\ref{stage1:3}).
			\State Compute the initial sum-rate $ c_{0} $, based on (\ref{WRF:1})$ \-- $(\ref{Capacity:1}). 
			\State $ c_{best} \gets c_{0} $
			\State $ \textbf{S}^{\star} \gets \textbf{S} $
			\State $\textbf{J}=[\textbf{j}_1,\textbf{j}_2,\cdots,\textbf{j}_K]$, where $\textbf{j}_k= [{1,2,\cdots,{N_{\text{r}}}}]^{\text{T}}.$
			\For {$k = 1$ to $K$}
			\State Find $ \hat{\textbf{j}}_k $ by sorting the elements of $ \textbf{j}_k $ in ascending order of $ \big\vert[\hat{\textbf{H}}]_{n,k}\big\vert$, $ n={1,2,\cdots,{N_{\text{r}}}} $.
			\EndFor 
			\State $ t = 0 $
			\For {$l = 1$ to $N_{\text{r}}-1$}
			\While {$ t < t_{max} $}
			\For {$k= 1$ to $K$} 
			\State $ i \gets [\hat{\textbf{J}}]_{l,k} $
			\State $ [{\textbf{S}}]_{i,k}= 0 $
			\EndFor 
			\State Calculate the sum-rate $ {c}_{l} $ based on (\ref{WRF:1}) $ \-- $ (\ref{Capacity:1}).
			\If {$ {c}_{l}>c_{best} $}
			\State Update $ c_{best} \gets c_{l} $, $ \textbf{S}^{\star} \gets \textbf{S} $.
			\State $ t \gets 0 $ 
			\Else 
			\State $ t \gets t + 1 $.
			\EndIf 
			\EndWhile 
			\EndFor \\			
			\Return $ {{\textbf{S}}^{\star}} $ 
		\end{algorithmic} 
		\label{alg:2}
	\end{algorithm}
	The overall procedure of the CM-ASS scheme with dynamic $ L $ is summarized in \cref{alg:2}. Steps 1$\--$4 are the same as those in DS-ASS. Steps 6$\--$8 construct a matrix $ \hat{\textbf{J}} $ comprising columns of the antenna indices, which are sorted in ascending order of the absolute values of entries of $ \hat{\textbf{H}} $. As shown in steps 13$\--$16, during the $ l $th iteration, one non-zero entry in each column of $ {\textbf{S}} $, corresponding to the antenna index in the $ l $th row of $ \hat{\textbf{J}} $, is converted to $ 0 $. Then, step 17 computes the corresponding sum-rate $ {c}_{l} $. Steps 18$ \-- $20 compare $ {c}_{l} $ to the current largest sum-rate $ {c}_{best} $. If $ {c}_{l} \geq {c}_{best} $, $ {c}_{best} $ and $ \textbf{S}^{\star}$ are updated to $ {c}_{l} $ and $ \textbf{S} $, respectively, and the search process is repeated. If $ {c}_{l}<{c}_{best}$ and $ t = t_{max} $, the process is terminated, and the current $ \textbf{S}^{\star}$ is adopted as the solution for the antenna subset selection.
	\subsubsection{CM-ASS with fixed L}	In this scheme, to further reduce the computational complexity, we set the number of active switches for each RF chain to a fixed value, i.e., $ {L_k}=L,\forall{_k} $. Therefore, in this scheme, each RF chain is connected to the same fixed number of active switches that correspond to $ L $ entries with the largest absolute values in each column of $ \hat{\textbf{H}}$. \par
	This scheme can be deduced from \cref{alg:2} by choosing and modifying certain steps. First, step 1 initializes $ \textbf{S}$ as $ \textbf{0}_{N_{\text{r}} \times K} $. This is followed by step 2 and steps 6$\--$9 of \cref{alg:2}. Finally, $ \textbf{S}^{\star}$ is generated by converting $ L $ zero entries in each column of $ {\textbf{S}} $ to $1$s, which correspond to the antenna indices in the first $ L $ rows of $ \hat{\textbf{J}} $. We note that the CM-ASS scheme with fixed $ L $ does not require the iterative process performed during steps 11$\--$25 of \cref{alg:2}, which can require up to $N_{\text{r}} -1$ iterations and the computation of sum-rate on step 17 in each iteration. Hence, we expect that CM-ASS with fixed $L$ requires substantially lower complexity than CM-ASS with dynamic $L$; we will verify this expectation through numerical results in Section V.
	\section{Energy Efficiency}\label{S: IV}	
	In this section, we compare the EE of the proposed architecture to other state-of-the-art MU-MIMO systems. The EE is defined as \cite {Rowell}
	\begin{equation} \label{ee}
	EE =\frac{R}{P_\text{T}}, 
	\end{equation}
	where $ R $ is the achievable sum-rate, and $ P_{\text{T}} $ is the total power consumption of the system.
	We adopt the power consumption model for the receiver in \cite{Rusu,Buzzi, Abbas} as the basis for comparison. The power consumed by a single low-noise amplifier (LNA) and two analog-to-digital converters (ADCs) for I and Q components is denoted by $ {P_{\text{LNA}}} $ and $ {P_{\text{ADC}}} $, respectively. The power consumed by a power splitter and power combiner is also denoted by $ {P_{\text{SP}}} $ and $ {P_{\text{C}}} $, respectively. Furthermore, the power consumed by a single switch, a VPS, and a CPS is denoted by $ P_{\text{SW}} $, $ P_{\text{VPS}} $, and $ P_{\text{CPS}} $, respectively, whereas the power consumed by an RF chain and a baseband signal processing block is represented by $ P_{\text{RFC}} $ and $ P_{\text{BB}} $, respectively. We note that $ P_{\text{RFC}} $ includes the power consumption of the mixer $ (P_{\text{M}}) $, local oscillator $(P_{\text{LO}}) $, low-pass filter $ (P_{\text{LPF}}) $, and base-band amplifier $ (P_{\text{BBamp}}) $; this power is given as \cite{Rusu}
	\begin{equation}
	P_{\text{RFC}} = P_{\text{M}}+P_{\text{LO}}+P_{\text{LPF}}+P_{\text{BBamp}}.		
	\end{equation}
	The total circuitry power consumption of the compared schemes can be expressed as follows:
	\begin{subequations}
		\begin{align}
		\begin{split}\label{FD}
	P_{\text{T}}^{\text{FD}}&=N_{\text{r}}(P_{\text{LNA}}+P_{\text{RFC}}+ P_{\text{ADC}})+P_{\text{BB}},		
		\end{split}\\
		\begin{split}\label{FVPS}
		P_{\text{T}}^{\text{FVPS}}&=N_{\text{r}}(P_{\text{LNA}}+P_{\text{SP}}+{K}P_{\text{VPS}})\\
		&+{K}(P_{\text{RFC}}+P_{\text{C}}+P_{\text{ADC}})+P_{\text{BB}},
		\end{split}\\
		\begin{split}\label{eeFCPS}
		P_{\text{T}}^{\text{FCPS}}&=N_{\text{r}}(P_{\text{LNA}}+P_{\text{SP}}+{K}P_{\text{SW}})+{K}(N_{\text{C}}P_{\text{CPS}}\\
		&+P_{\text{C}}(N_{\text{C}}+1)+P_{\text{RFC}}+P_{\text{ADC}})+P_{\text{BB}},
		\end{split}\\
		\begin{split}\label{eeDS-ASS}
		P_{\text{T}}^{\text{Prop}}&=N_{\text{r}}(P_{\text{LNA}}+P_{\text{SP}})+ \sum\limits _{k=1}^{K}L_{k}P_{\text{SW}}+{K}(N_{\text{C}}P_{\text{CPS}}\\
		&+P_{\text{C}}(N_{\text{C}}+1)+P_{\text{RFC}}+P_{\text{ADC}})+P_{\text{BB}},
		\end{split}
		\end{align}
	\end{subequations}
	where $ {P_{\text{T}}^{\text{FD}}} $, $ {P_{\text{T}}^{\text{FVPS}}} $, and $ {P_{\text{T}}^{\text{FCPS}}} $ indicate the total power consumptions of FD, fully connected VPS (FVPS), and FCPS architectures \cite{Abbas,Buzzi,Rusu}, respectively. Furthermore, $ {P_{\text{T}}^{\text{Prop}}} $ represents the total power consumption of the proposed architecture. In the case of CM-ASS, we have $ {L_{k} = L}$, $ k=1,\cdots, {K}$, and hence (\ref{eeDS-ASS}) can be rewritten as
	\begin{equation}\label{eeCM-ASS}
	\begin{split}
	P_{\text{T}}^{\text{Prop}}&=N_{\text{r}}(P_{\text{LNA}}+P_{\text{SP}})+{K}({L}P_{\text{SW}}+N_{\text{C}}P_{\text{CPS}}\\
	&+P_{\text{C}}(N_{\text{C}}+1)+P_{\text{RFC}}+P_{\text{ADC}})+P_{\text{BB}}.
	\end{split}
	\end{equation}
	In \cref{tab:table1}, the assumed power consumption of each component is presented based on the assumptions made in recent studies of EE analysis for a reference carrier frequency of $ f_{c} = 60$ GHz.\par
	\bgroup
		\def\arraystretch{1.5}%
		\begin{table} [!h]
			\centering
			\begin{center}
				\caption{Power consumption of each component in the receiver}
				\label{tab:table1}
				\begin{tabular}{ c | c | c }
					\hline
					\textbf{Hardware component} & \textbf{Notation} & \textbf{Power consumption} \\ \hline
					\hline
					Low noise amplifier \cite{Rusu} & $ P_{\text{LNA}} $ & 20 mW \\ \hline
					Variable phase shifter \cite{Rusu,Graauw} & $ P_{\text{VPS}} $ & 30 mW \\ \hline
					Combiner \cite{Graauw} & $ P_{\text{C}} $ & 19.5 mW \\ \hline
					Splitter \cite{Graauw} & $ P_{\text{SP}} $ & 19.5 mW \\ \hline
					Switch \cite{Rusu} & $ P_{\text{SW}}$ & 5 mW \\ \hline
					Constant phase shifter \cite{Rusu} & $ P_{\text{CPS}}$ & 5 mW \\ \hline
					RF chain \cite{Rusu} & $ P_{\text{RFC}} $ & 40 mW \\ \hline
					Baseband processor\cite{Rusu} & $ P_{\text{BB}} $ & 200 mW \\ \hline
					ADC \cite{Rusu} & {$ P_{\text{ADC}} $} & {200 mW}\\ \hline
				\end{tabular}
			\end{center}
		\end{table}
	\egroup
	Based on (\ref{eeDS-ASS}), (\ref{eeCM-ASS}), and \cref{tab:table1}, the proposed structure is expected to have less power consumption in the RF circuit as compared with the FD and FVPS schemes because it employs low-power components, i.e., CPSs and switches. Furthermore, owing to $ L_{k} \leq N_{\text{r}}$ and $ L \leq N_{\text{r}}$ in (\ref{eeDS-ASS}) and (\ref{eeCM-ASS}), respectively, the power consumption in the RF circuit of the proposed structure is lower than that of FCPS in (\ref{eeFCPS}). Specifically, inactive switches do not consume power, reducing overall power consumption, whereas, interestingly, setting a subset of switches in inactive states can also enhance the sum-rate. Related numerical simulation results will be provided in the next section. \par 
	\bgroup
		\def\arraystretch{1.5}
		\begin{table} [htbp]	
			\centering
			\caption{Comparison of the power consumption of different architectures for $ K = {\{ 4,16\}} $, $ N_{\text{C}} = 8 $, and $ N_{\text{r}} ={\{64,128\}} $}
			\label{tab:table2}
			\begin{tabular}{c|c|c|c|c|c}
				\hline
				\multicolumn{2}{c|}{\multirow{2}{*}{Algorithm}}&
				\multicolumn{2}{c|}{$ N_{\text{r}} = 64 $} & \multicolumn{2}{c}{$ N_{\text{r}} = 128 $}\\ \cline{3-6} 
				\multicolumn2{c|}{}& {$K = 4 $} & {$ K = 16$} & {$K = 4 $} & {$ K =16$} \\ \cline{3-6} 
				\hline	\hline				
				\multicolumn{2}{c|}{${P_{\text{T}}^{\text{FD}}}$} & {16.84 W}& {16.84 W} & {33.48 W}& {33.48 W}\\ \hline
				\multicolumn{2}{c|}{$ {P_{\text{T}}^{\text{FVPS}}}$}& {11.45 W}& {37.60 W}& {21.65 W}& {70.85 W} \\ \hline
				\multicolumn{2}{c|}{$ {P_{\text{T}}^{\text{FCPS}}}$} & {5.83 W} & {15.14 W}& {9.64 W}& {22.78 W} \\ \hline
				\multirow {2}{*}{$ P_{\text{T}}^{\text{Prop}} $}
				&{$ L = {0.5}{N_{\text{r}}}$} & {5.19 W} & {12.58 W}& {8.36 W}& {17.66 W} \\ \cline{2-6} 
				&{$ L = {0.75}{N_{\text{r}}} $} & {5.51 W}& {13.86 W}& {9.00 W}& {20.22 W} \\ \cline{2-6} 	     
				\hline		 		
			\end{tabular}
		\end{table}
	\egroup
	\cref{tab:table2} shows a comparison of the power consumption for various architectures. For the proposed architecture, we assume (\ref{eeCM-ASS}), where $ L_{k} = L, \forall k$. In \cref{tab:table2}, we observe that proposed architecture has the lowest power consumption among those compared, owing to the use of active/inactive switches and low-power CPSs. For example, the proposed scheme with $ L = 0.5 N_{\text{r}}$ can achieve power-reduction ratios in the ranges of $25.3\%\--75.0\%$, $54.7\%\--75.1\%$, and $11.0\%\--22.5\%$ over the FD, FVPS, and FCPS architectures, respectively.
	 		
	\section{Simulation Results}\label{S: V}
	In this section, we present numerical simulation results to evaluate the performance of the proposed schemes. We compare the proposed schemes with the Gram--Schmidt-based algorithm in \cite{Xiao} for FVPS, the improved quasi-coherent combining algorithm in \cite{Sah} for FCPS, and the MMSE receiver proposed in \cite{Bengtsson} for FD schemes. In the simulation results, we consider an environment with $ N_{\text{p}} = 15 $ propagation paths between each single-antenna user and the BS {\cite{Sohrabi}, uniformly distributed random AoAs within [$ 0,2\pi $], and $ d=\frac{\lambda}{2} $, unless otherwise stated. We assume that the number of RF chains is equal to the number of users, i.e., $ N_{\text{RF}} = K$.
		
		\subsection{Simulation Results for Spectral Efficiency}
		In this subsection, we present various simulation results for the SE performances of the proposed schemes and other conventional schemes in different environments.\par
		\cref{fig:3} plots the SE for various values of $L$ in the CM-ASS scheme with fixed $L$ for a system with $ N_{\text{r}} = 64 $, $ K=16 $, $ N_{{\text{C}}}= 8 $, and SNR $ = 0 $ dB. From \cref{fig:3}, we observe that maximum SE is achieved at $ L\approx0.75N_{\text{r}} $, i.e., approximately $ 25\% $ of switches per RF chain are inactive. In \cref{fig:3}, we also observe two intersection points between the FCPS and CM-ASS schemes: (i) at $ L\approx0.5N_{\text{r}} $, i.e., approximately $ 50\% $ of switches per RF chain are active; and (ii) at $ L = N_{\text{r}} $, i.e., all switches are active. These results imply that the CM-ASS scheme can achieve the performance of FCPS only when $ L\approx0.5N_{\text{r}} $ switches are active for each RF chain. They also numerically justify that the exclusion of some antennas from signal combining can improve the performance. Similar results can be observed in other simulations performed under different environments and SNRs. Based on these observations, in the remaining simulation results, $L$ for the CM-ASS scheme with fixed $L$ is set to $ 0.5N_{\text{r}} $ and $ 0.75N_{\text{r}} $. This is because the CM-ASS scheme with fixed $L=0.75N_{\text{r}}$ achieves near-optimal performance, and the same scheme with $ L=0.5N_{\text{r}} $ achieves performance comparable to that of the FCPS while consuming much less power.\par
		\begin{figure}[t]
			\centering
			\includegraphics[scale = 0.57]{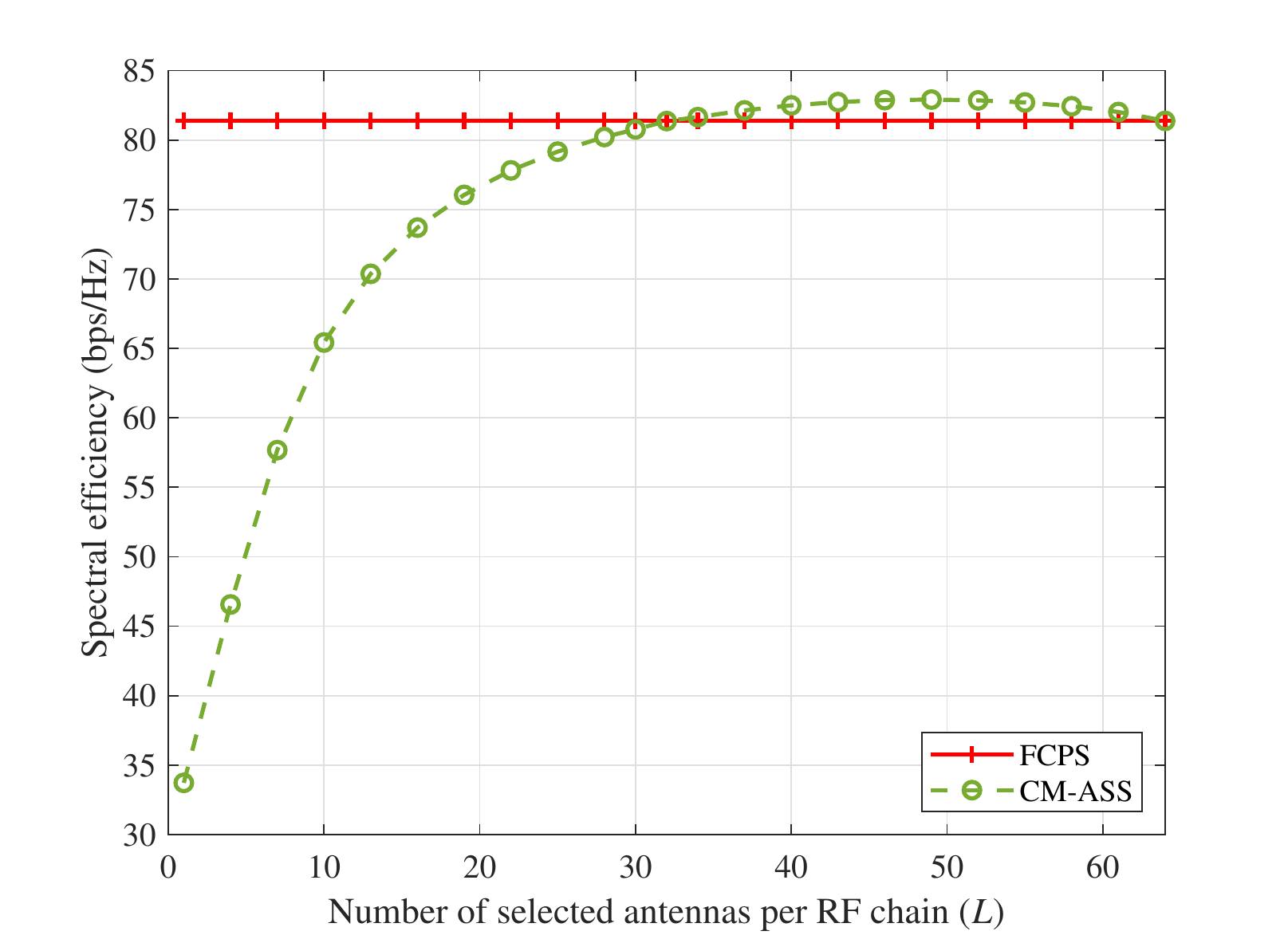}
			\caption{SE versus the number of selected antennas per RF chain for a system with $ N_{\text{r}} = 64 $, $ K = 16 $, $ N_{{\text{C}}}= 8 $, and SNR $ = 0 $ dB.}
			\label{fig:3}
		\end{figure}
		\begin{figure}[t]
			\centering
			\includegraphics[scale = 0.57]{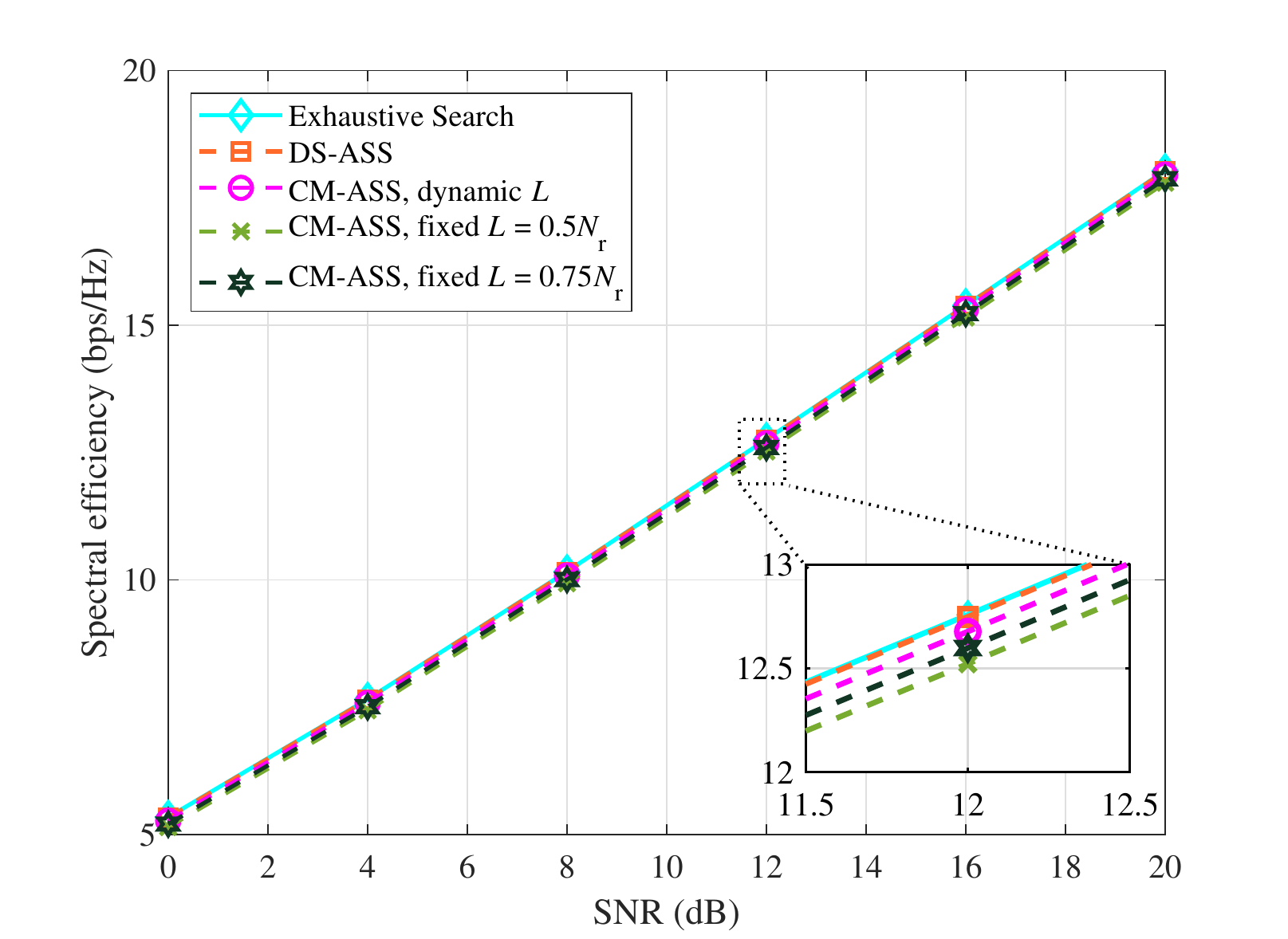}
			\caption{SE versus SNR for a system with $ N_{\text{r}} = 8 $, $ K=2 $, and $ N_{{\text{C}}}=8 $.}
			\label{fig:4}
		\end{figure}
		\cref{fig:4} compares the SE achieved by the proposed low-complexity antenna subset selection schemes with that achieved by the exhaustive search-based antenna subset selection. Owing to an extremely large number of possible combinations that must be examined in the exhaustive search for large $ N_{\text{r}} $ and $K$, a relatively small system with $ N_{\text{r}} = 8 $, $ K=2 $, and $ N_{\text{C}} = 8 $ is considered for this scenario. In \cref{fig:4}, it is observed that the proposed DS-ASS achieves almost the same SE as the exhaustive search-based antenna subset selection. \cref{fig:4} also shows that CM-ASS schemes achieve SE performances comparable to those of the exhaustive search-based antenna subset selection scheme. In particular, the performance loss of CM-ASS with dynamic $ L $ with respect to exhaustive search is only $ 0.6\% $ at SNR $ =12 $ dB. \par
		\begin{figure}[t]
			\centering
			\includegraphics[scale = 0.57]{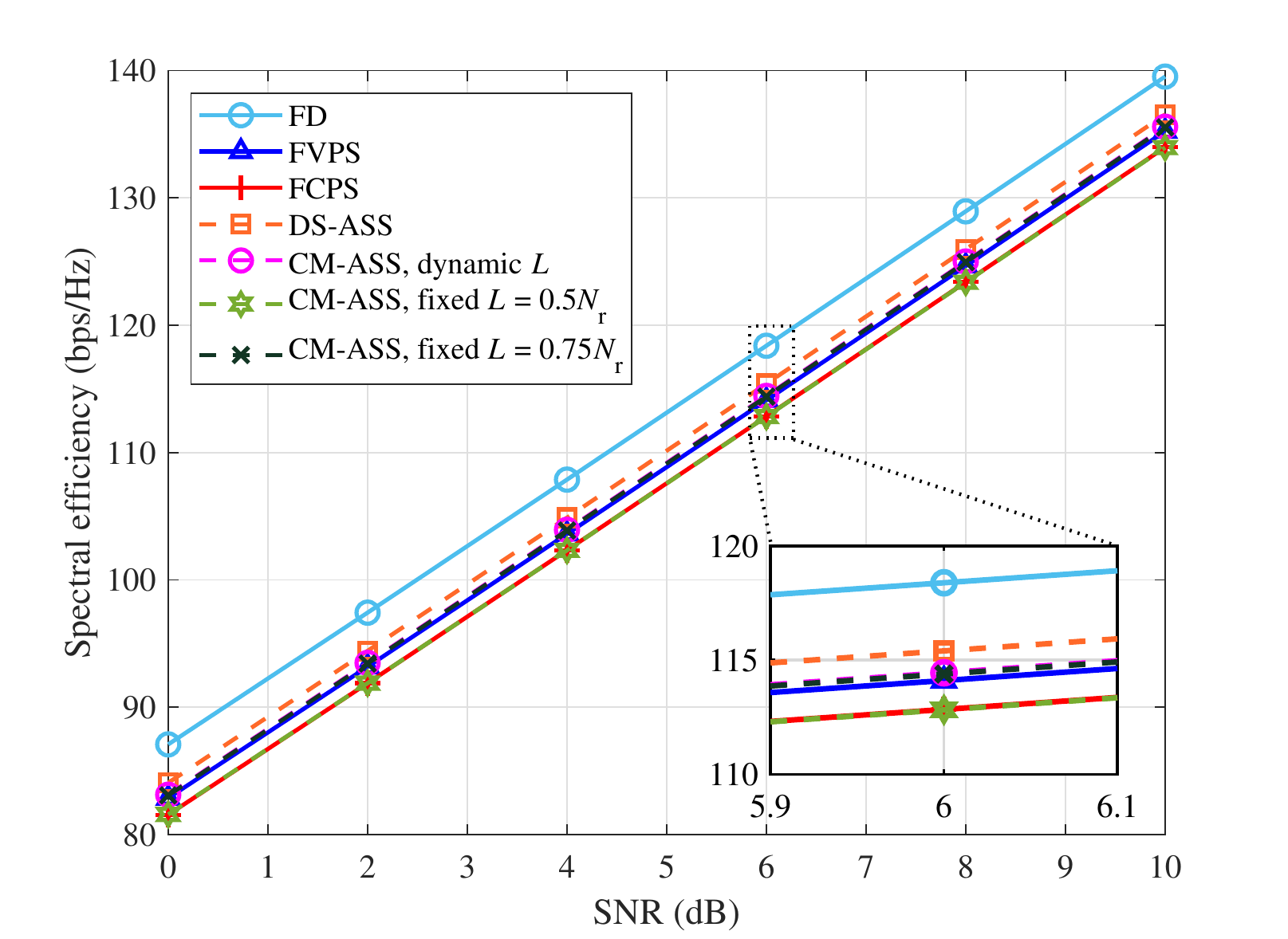}
			\caption{SE versus SNR for a system with $ N_{\text{r}} = 64 $, $ K = 16 $, and $ N_{\text{C}} = 8$.} 
			\label{fig:5} 
		\end{figure}
		In \cref{fig:5}, the SE performances of various architectures are presented for $ N_{\text{r}} = 64 $, $ K = 16$, and $ N_{\text{C}} =8 $. This result demonstrates that the proposed schemes outperform both FCPS and FVPS schemes. Note that in \cref{fig:5}, both the proposed architecture and the FCPS scheme employ $ N_{\text{C}}\times K = 128 $ CPSs, whereas the FVPS scheme employs $ N_{\text{r}}\times K = 1024 $ VPSs. However, in the proposed schemes, we perform antenna subset selection, which implies that we use smaller numbers of active switches per RF chain as compared with the FCPS architecture, where all switches are active. The proposed schemes achieve improved performances with respect to the FCPS and FVPS schemes because, for each RF chain, the subset of received signals, which can lower sum-rates, is excluded from signal combining through antenna selection.\par
		
		\begin{figure}[t]
			\centering
			\includegraphics[scale = 0.57]{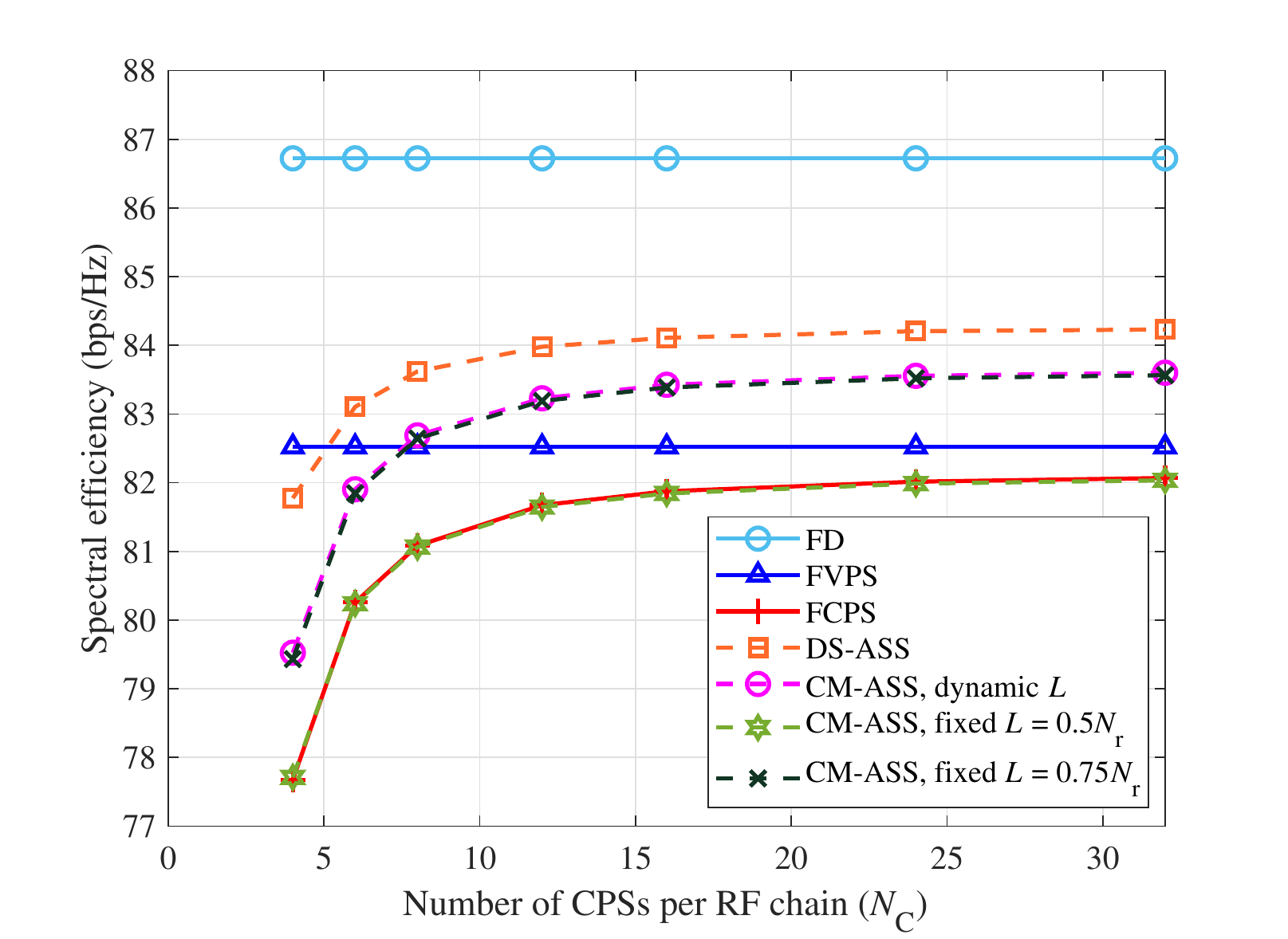}
			\caption{SE versus $ N_{\text{C}} $ for a system with $ N_{\text{r}} = 64 $, $ K = 16 $, and SNR = 0 dB.}
			\label{fig:6}
		\end{figure}
		\begin{figure}[t]
			\centering
			\includegraphics[scale = 0.45]{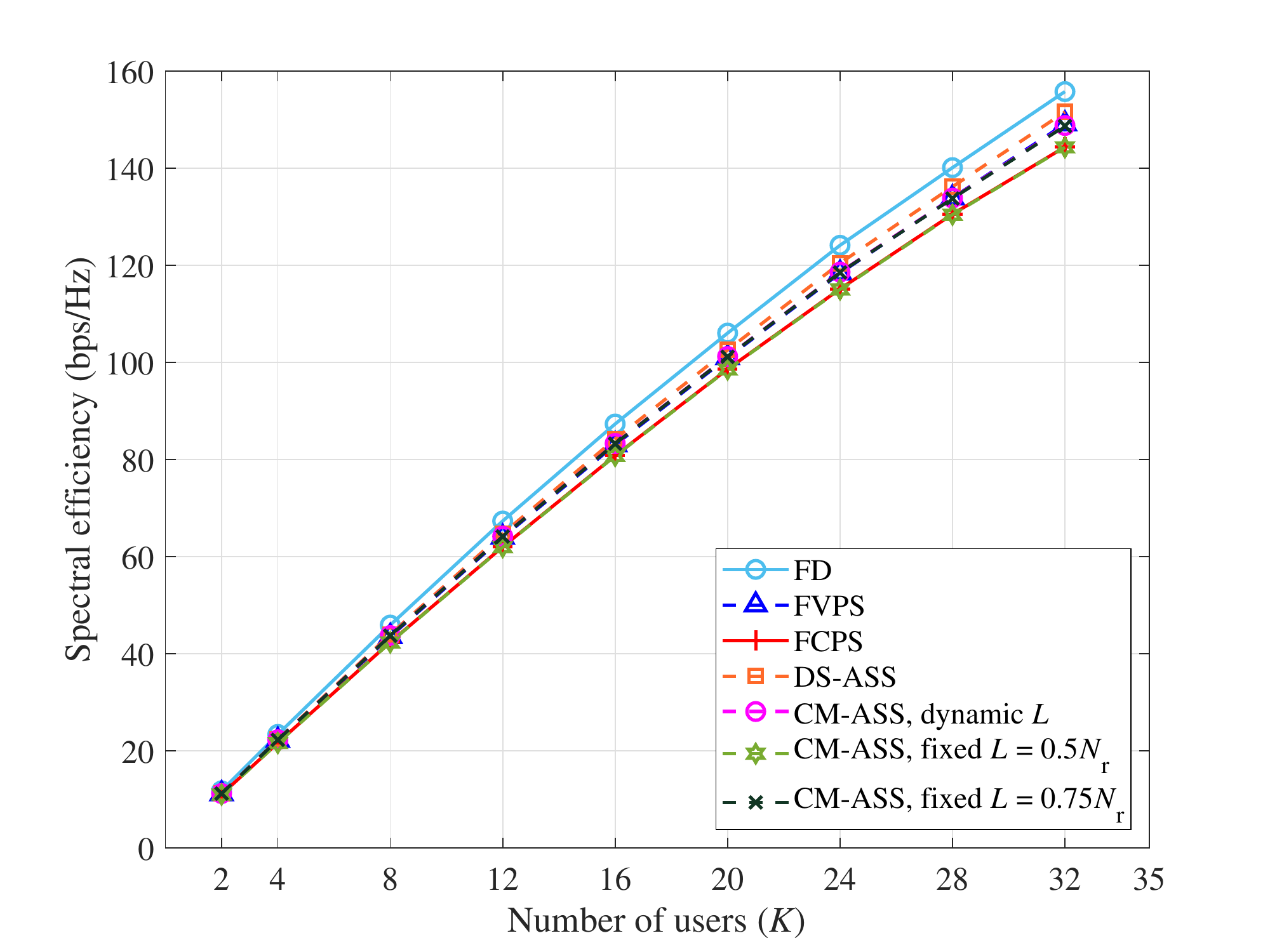}
			\caption{SE versus the number of users for a system with $ N_{\text{r}} = 64 $, $ N_{\text{C}} = 8 $, and SNR $ = 0 $ dB.}
			\label{fig:7}
		\end{figure}
		\cref{fig:6} shows the SE versus the number of CPSs per RF chain, $ N_{\text{C}} = {\{4,6,8,12,16,24,32}\}$, for a system with $ N_{\text{r}} = 64 $, and $ K = N_{\text{RF}} = 16$. \cref{fig:6} shows that the performances of both the proposed schemes and the FCPS scheme improve with $ N_{\text{C}}$. However, the FCPS scheme's performance does not exceed that of the FVPS scheme even for large $ N_{\text{C}}$. In contrast, eight CPSs per RF chain are enough for the proposed schemes to outperform the FVPS scheme. Another observation from \cref{fig:6} is that the performance improvement is almost negligible after $N_{\text{C}} = 16$ for the proposed structure as well as for the FCPS scheme, implying that $N_{\text{C}} = 16$ provides a sufficient level of granularity for phase quantization.\par	
		\cref{fig:7} presents the SE when the number of users varies in a system with $ N_{\text{r}} = 64 $ and $ N_{\text{C}} = 8 $. \cref{fig:7} shows that the proposed schemes outperform the FCPS throughout the entire range of $ K $. In particular, the gain of DS-ASS over the FCPS scheme reaches approximately $ 6 \% $ when $ K $ is large. As the number of users increases, the system becomes more interference-limited. Hence, in an environment with large $ K $, exclusion of received signals that contribute more to the interference power than to the desired signal's power, which is executed by setting switches to inactive states, can provide higher performance gains, as shown in \cref{fig:7}.
		\subsection{Simulation Results for Energy Efficiency} \label{EE-Section}
		In this section, we compare the EE of the proposed structure with that of conventional schemes, according to (\ref{ee}). \par
		\begin{figure}[t]
			\centering
			\includegraphics[scale = 0.57]{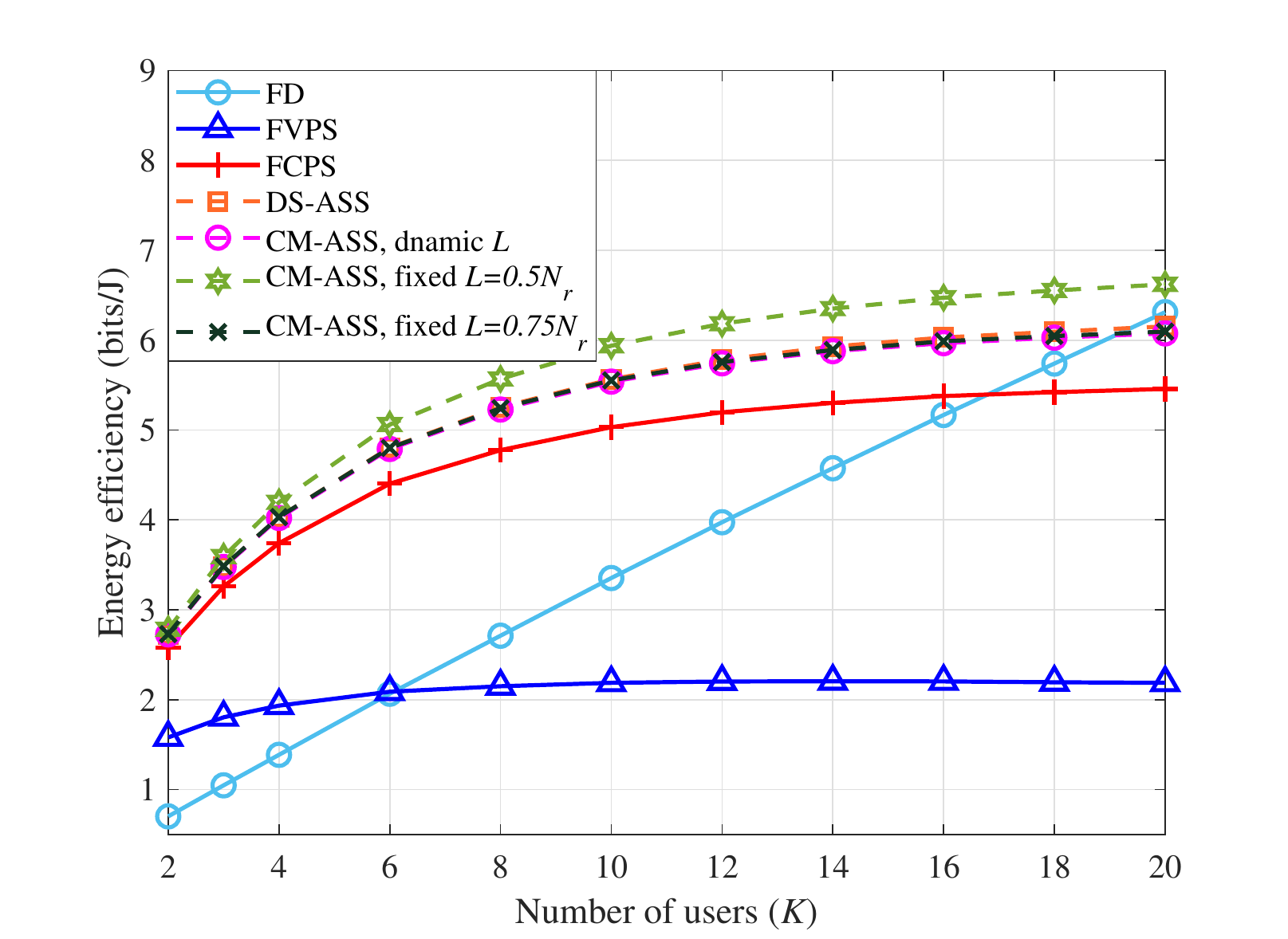}
			\caption{EE versus the number of users for a system with $ N_{\text{r}} = 64 $, $ N_{\text{C}} = 8 $, and SNR $ = 0 $ dB.}
			\label{fig:8}
		\end{figure}
		\cref{fig:8} shows the impact of the number of users on EE. In \cref{fig:8}, we observe that for small and moderate numbers of users, the proposed schemes offer remarkably improved EE with respect to conventional schemes. Although the FCPS and proposed schemes enjoy higher EE than the FD and FVPS schemes for small and moderate numbers of users, \cref{fig:8} shows clear gains in the EE offered by the proposed schemes over the FCPS scheme. For example, for $K = 8$, the EE gains of the CM-ASS scheme with fixed $L = 0.5N_{\text{r}}$ over the FD, FVPS, and FCPS schemes are approximately $106.5\%$, $160.5\%$, and $19.2\%$, respectively, and the gains of the DS-ASS scheme over the FD, FVPS, and FCPS schemes are approximately $95.4\% $, $146.5\%$, and $12.8\%$, respectively. \par
		In \cref{fig:8}, we also observe that the CM-ASS scheme with fixed $ L = 0.5N_{\text{r}}$ achieves higher EE than the other proposed schemes, namely, DS-ASS and CM-ASS with dynamic $L$. This is because DS-ASS and CM-ASS with dynamic $L$ are optimized to enhance the SE, rather than maximizing the EE. In contrast, in the CM-ASS with fixed $ L = 0.5N_{\text{r}}$, only half the switches are activated at any given time; hence, its power consumption can be lower than that of the other schemes, thereby yielding higher EE at the cost of a moderate reduction in SE.\par 		
		\begin{figure}[t]
			\centering
			\includegraphics[scale = 0.57]{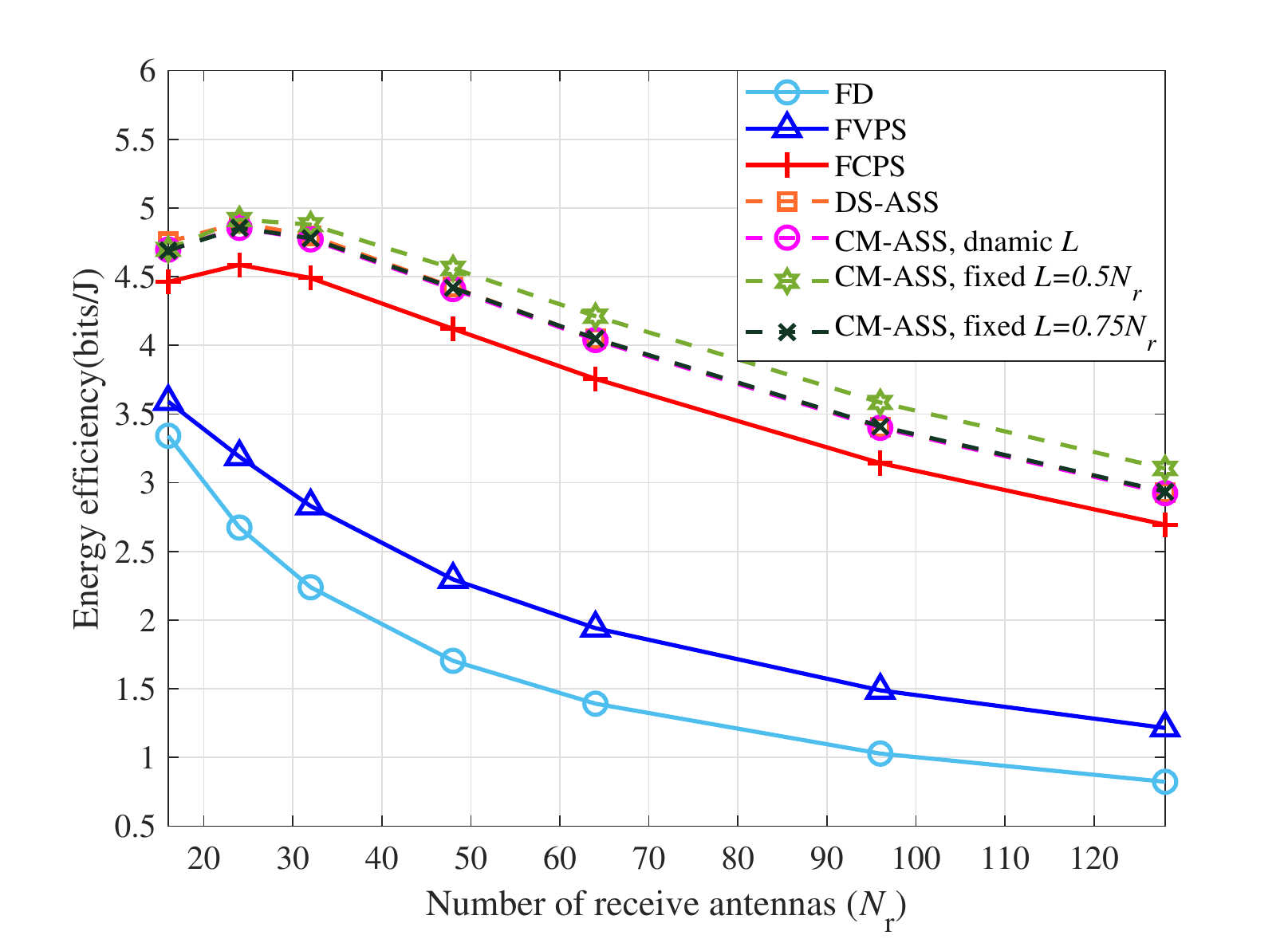}
			\caption{EE versus the number of receive antennas for a system with $K = 4$, $N_{\text{C}} = 8$, and SNR $ = 0 $ dB.}
			\label{fig:9.1}
		\end{figure}		\begin{figure}[t]
			\centering
			\includegraphics[scale = 0.57]{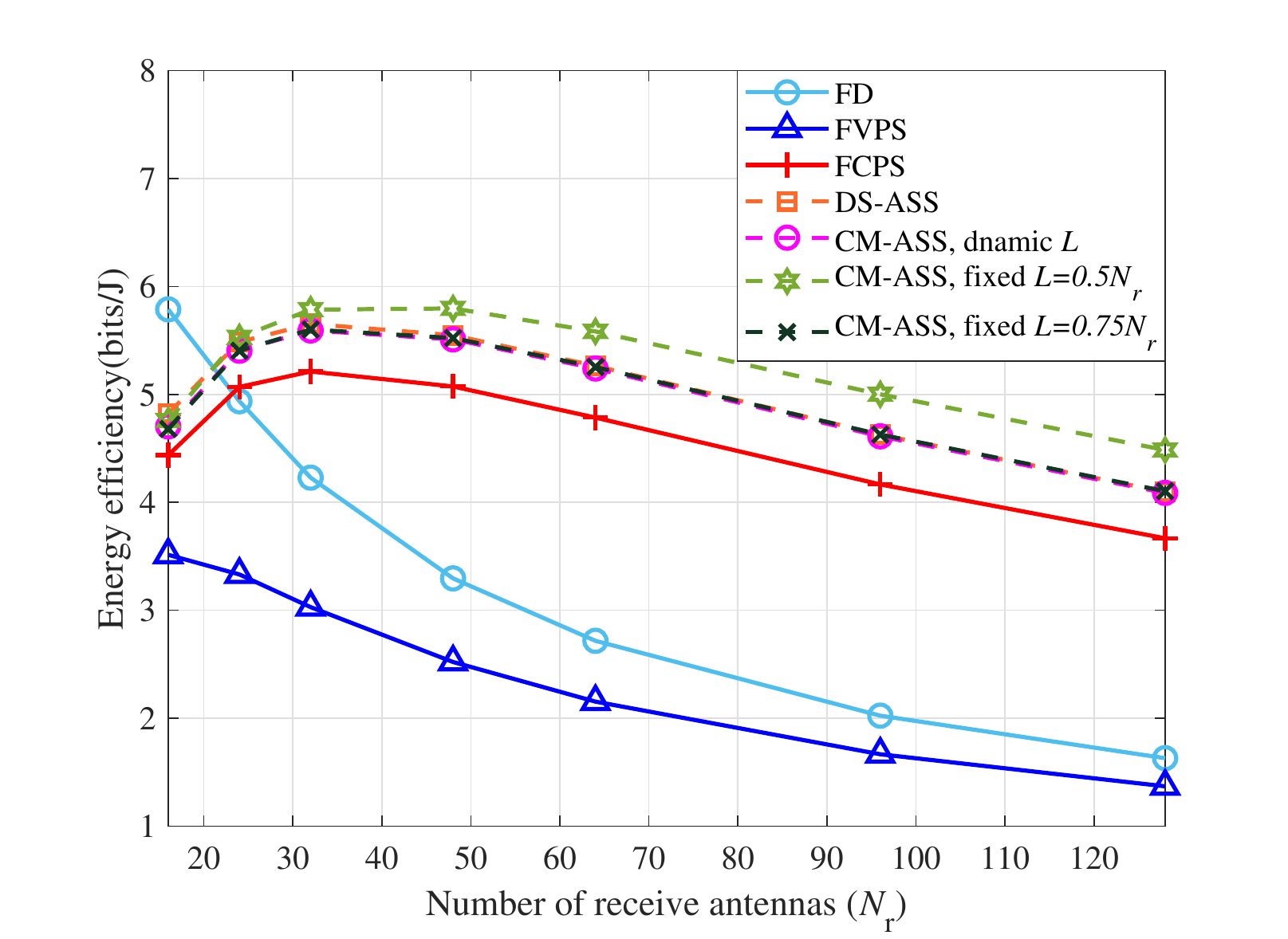}
			\caption{EE versus the number of receive antennas for a system with $ K = 8 $, $ N_{\text{C}} = 8 $, and SNR $ = 0 $ dB.}
			\label{fig:9.2}
		\end{figure}	
		\cref{fig:9.1} and \cref{fig:9.2} illustrate the EE for various numbers of receive antennas. In \cref{fig:9.1} and \cref{fig:9.2}, we consider systems with $ K = N_{\text{RF}} = 4$ and $ K = N_{\text{RF}} = 8$, respectively. Both figures show that the proposed schemes achieve higher EE than conventional schemes in almost the entire region. For example, in \cref{fig:9.1}, for a small number of antennas such as $ N_{\text{r}} = 16 $, the proposes schemes have nearly the same EE, and their performance gains over the FD, FVPS, and FCPS schemes are approximately $ 43.6\% $, $ 33.3\% $, and $ 9.1\% $, respectively. In addition, in \cref{fig:9.1}, for $ N_{\text{r}} = 128 $, the EE gains of the CM-ASS scheme with fixed $ L = 0.5N_{\text{r}}$ over the FD, FVPS, and FCPS schemes are $ 277.9\% $, $ 156.0\% $, and $ 17.4\% $, respectively, and the DS-ASS, CM-ASS with dynamic $ L $, and CM-ASS with fixed $ L = 0.75N_{\text{r}}$ schemes achieve nearly the same EE gains of approximately $ 258.4\% $, $ 142.9\% $, and $ 11.3\% $ over the FD, FVPS, and FCPS schemes, respectively. Similar to \cref{fig:9.1}, in \cref{fig:9.2}, the EE performance gains of the proposed schemes over conventional schemes are clear. \par
		Finally, the complexity of the proposed schemes should be analyzed. This complexity analysis is performed by counting the numbers of complex floating-point operations \cite{Golub}. The computational complexities of the proposed schemes mainly increase with the number of iterations required to perform per-RF chain antenna subset selection. The search process to obtain a solution of antenna subset selection requires $ O(i(N_{\text{r}}K^2+K^3)) $ and $ O(i(N_{\text{r}}K^2+N_{\text{r}}K)) $ computational complexity for digital MMSE combining (\ref{WBB1}) and the computation of the sum-rates in (\ref{Capacity:1}), respectively, where $ i $ is the maximum iteration number. For the DS-ASS scheme, we have $i=\sum_{j=K+1}^{N_{\text{r}}{K}}{j} $, whereas for the CM-ASS scheme with dynamic $ L $, $ i= N_{\text{r}} $, and for the CM-ASS scheme with fixed $ L $, there is no required search process, which implies $ i = 1 $. Note that this is a huge reduction in the maximum iteration number as compared to an exhaustive search, which requires $ 2^{N_{\text{r}}{K}}$ iterations to test all possible combinations. Furthermore, because the proposed schemes apply early termination, the actual number of iterations required to obtain a solution of antenna subset selection can be reduced. The other computational loads are obtained from $ O(N_{\text{r}}K^2) $ operations for QR decomposition to generate $ \hat{\textbf{H}} $ \cite{Sah} and $ O(N_{\text{r}}KN_{\text{C}}) $ operations in (\ref{stage1:1}) $ \-- $ (\ref{stage1:3}) to obtain ${\tilde{\textbf{W}}_{\text{RF}}}$. \par
		\cref{fig:10} compares the number of the complex floating-point operations for a system with $ N_{\text{RF}} = 4 $ and $ N_{\text{C}} = 8 $. In \cref{fig:10}, the proposed schemes are observed to require significantly lower complexity than that required by an exhaustive search. This shows that the complexity of exhaustive search-based antenna subset selection increases very rapidly with the number of receive antennas as compared with the proposed schemes. In \cref{fig:10}, we also observe that the DS-ASS scheme has higher complexity than the CM-ASS schemes because the DS-ASS scheme performs more iterations to obtain the solution of antenna subset selection. In contrast, CM-ASS schemes with fixed $ L $ have the lowest complexity because only one combination is chosen as the candidate for the antenna subset selection solution. Specifically, CM-ASS with fixed $ L $ requires $ 99.9\% $ and $ 99.7\% $ lower computational loads compared to DS-ASS and CM-ASS with dynamic $ L $ for $ N_{\text{r}} = 32 $.
		\begin{figure}[t]
			\centering
			\includegraphics[scale = 0.57]{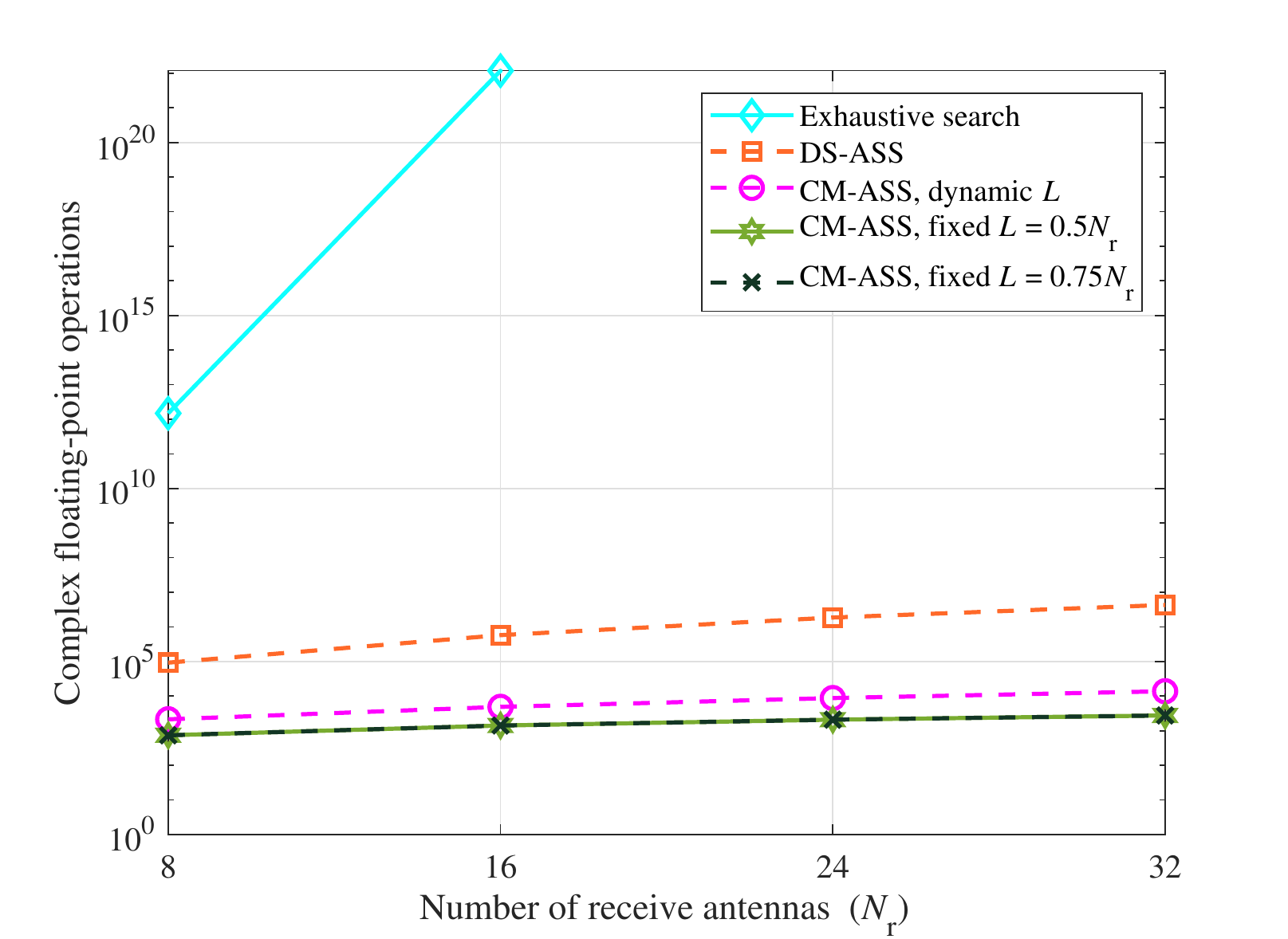}
			\caption{Number of complex floating-point operations versus the number of receive antennas for a system with $ N_{\text{RF}} = K= 4 $, $ N_{\text{C}}= 8 $, and SNR $ = 0 $ dB.}
			\label{fig:10}
		\end{figure}
		\section{Conclusion}\label{S: VI}
		In this study, we have presented a novel hybrid combining architecture for mmWave uplink MU-MIMO systems, in which per-RF chain antenna subset selection is exploited. In the proposed architecture, by deactivating some switches, the power consumption of the RF circuit can be reduced while simultaneously enhancing the sum-rate. We have developed low-complexity algorithms to perform per-RF chain antenna subset selection. The numerical simulation results show that the proposed structure can provide both higher EE and higher SE than conventional FVPS and FCPS hybrid beamforming schemes. Specifically, the proposed CM-ASS with fixed $L$ achieves EE performance gains of $ 9.1 \% \-- 19.2\% $ over the FCPS scheme, whereas its gains over the FVPS scheme are $ 33.3\% \-- 160.5\% $, and its gains over the FD scheme are $ 43.3\% \-- 277.9\% $. One possible extension of the current work would be the design of hybrid combiners using double phase shifters for each antenna \cite{Huang,Kung} to further improve the sum-rate performance of the hybrid combining architecture based on antenna subset selection. Furthermore, the proposed scheme is a potential candidate for the vehicular communication infrastructure because of its low power requirement and improved SE performance. It would be interesting to extend this study to multipath fast-fading channels in vehicular communication and evaluate the performance.} 
		\ifCLASSOPTIONcaptionsoff
		\newpage
		\fi
		\bibliographystyle{IEEEtran}
		\bibliography{references}	

\begin{thebibliography}{10}
\providecommand{\url}[1]{#1}
\csname url@samestyle\endcsname
\providecommand{\newblock}{\relax}
\providecommand{\bibinfo}[2]{#2}
\providecommand{\BIBentrySTDinterwordspacing}{\spaceskip=0pt\relax}
\providecommand{\BIBentryALTinterwordstretchfactor}{4}
\providecommand{\BIBentryALTinterwordspacing}{\spaceskip=\fontdimen2\font plus
\BIBentryALTinterwordstretchfactor\fontdimen3\font minus
  \fontdimen4\font\relax}
\providecommand{\BIBforeignlanguage}[2]{{%
\expandafter\ifx\csname l@#1\endcsname\relax
\typeout{** WARNING: IEEEtran.bst: No hyphenation pattern has been}%
\typeout{** loaded for the language `#1'. Using the pattern for}%
\typeout{** the default language instead.}%
\else
\language=\csname l@#1\endcsname
\fi
#2}}
\providecommand{\BIBdecl}{\relax}
\BIBdecl

\bibitem{Khan1}
Z.~{Pi} and F.~{Khan}, ``A millimeter-wave massive {MIMO} system for next
  generation mobile broadband,'' in \emph{Proc. Asilomar Conf. Signal Syst.
  Comput.}, Nov. 2012, pp. 693--698.

\bibitem{Rapp}
T.~S. {Rappaport}, S.~{Sun}, R.~{Mayzus}, H.~{Zhao}, Y.~{Azar}, K.~{Wang},
  G.~N. {Wong}, J.~K. {Schulz}, M.~{Samimi}, and F.~{Gutierrez}, ``Millimeter
  wave mobile communications for {5G} cellular: It will work!'' \emph{IEEE
  Access}, vol.~1, pp. 335--349, May 2013.

\bibitem{Sun}
S.~{Sun}, T.~S. {Rappaport}, R.~W. {Heath}, A.~{Nix}, and S.~{Rangan}, ``{MIMO}
  for millimeter-wave wireless communications: Beamforming, spatial
  multiplexing, or both?'' \emph{IEEE Commun. Mag.}, vol.~52, no.~12, pp.
  110--121, Dec. 2014.

\bibitem{Ayanoglu}
A.~L. {Swindlehurst}, E.~{Ayanoglu}, P.~{Heydari}, and F.~{Capolino},
  ``Millimeter-wave massive {MIMO}: The next wireless revolution?'' \emph{IEEE
  Commun. Mag.}, vol.~52, no.~9, pp. 56--62, Sep. 2014.

\bibitem{Larsson}
E.~G. {Larsson}, O.~{Edfors}, F.~{Tufvesson}, and T.~L. {Marzetta}, ``Massive
  {MIMO} for next generation wireless systems,'' \emph{IEEE Commun. Mag.},
  vol.~52, no.~2, pp. 186--195, Feb. 2014.

\bibitem{Doan}
C.~H. {Doan}, S.~{Emami}, D.~A. {Sobel}, A.~M. {Niknejad}, and R.~W.
  {Brodersen}, ``Design considerations for 60 {GHz} {CMOS} radios,'' \emph{IEEE
  Commun. Mag.}, vol.~42, no.~12, pp. 132--140, Dec. 2004.

\bibitem{Zhang}
S.~{Zhang}, C.~{Guo}, T.~{Wang}, and W.~{Zhang}, ``On–off analog beamforming
  for massive {MIMO},'' \emph{IEEE Trans. Veh. Technol.}, vol.~67, no.~5, pp.
  4113--4123, May 2018.

\bibitem{Rusu}
R.~{Méndez-Rial}, C.~{Rusu}, N.~{González-Prelcic}, A.~{Alkhateeb}, and R.~W.
  {Heath}, ``Hybrid {MIMO} architectures for millimeter wave communications:
  Phase shifters or switches?'' \emph{IEEE Access}, vol.~4, pp. 247--267, 2016.

\bibitem{Sohrabi}
F.~{Sohrabi} and W.~{Yu}, ``Hybrid digital and analog beamforming design for
  large-scale antenna arrays,'' \emph{IEEE J. Sel. Topics Signal Process.},
  vol.~10, no.~3, pp. 501--513, Apr. 2016.

\bibitem{Ayach}
O.~E. {Ayach}, S.~{Rajagopal}, S.~{Abu-Surra}, Z.~{Pi}, and R.~W. {Heath},
  ``Spatially sparse precoding in millimeter wave {MIMO} systems,'' \emph{IEEE
  Trans. Wireless Commun.}, vol.~13, no.~3, pp. 1499--1513, Mar. 2014.

\bibitem{Rowell}
S.~{Han}, C.~{I}, Z.~{Xu}, and C.~{Rowell}, ``Large-scale antenna systems with
  hybrid analog and digital beamforming for millimeter wave {5G},'' \emph{IEEE
  Commun. Mag.}, vol.~53, no.~1, pp. 186--194, Jan. 2015.

\bibitem{Heath}
A.~{Alkhateeb}, G.~{Leus}, and R.~W. {Heath}, ``Limited feedback hybrid
  precoding for multi-user millimeter wave systems,'' \emph{IEEE Trans.
  Wireless Commun.}, vol.~14, no.~11, pp. 6481--6494, Nov. 2015.

\bibitem{Abbas}
W.~B. {Abbas}, F.~{Gomez-Cuba}, and M.~{Zorzi}, ``Millimeter wave receiver
  efficiency: A comprehensive comparison of beamforming schemes with low
  resolution {ADCs},'' \emph{IEEE Trans. Commun.}, vol.~16, no.~12, pp.
  8131--8146, Dec. 2017.

\bibitem{Zhu}
X.~{Zhu}, Z.~{Wang}, L.~{Dai}, and Q.~{Wang}, ``Adaptive hybrid precoding for
  multi-user massive {MIMO},'' \emph{IEEE Commun. Let.}, vol.~20, no.~4, pp.
  776--779, Apr. 2016.

\bibitem{Masouros}
A.~{Li} and C.~{Masouros}, ``Hybrid precoding and combining design for
  millimeter-wave multi-user {MIMO} based on {SVD},'' in \emph{2017 IEEE Int.
  Conf. Commun. (ICC)}, May 2017, pp. 1--6.

\bibitem{Graauw}
Y.~{Yu}, P.~G.~M. {Baltus}, A.~{de Graauw}, E.~{van der Heijden}, C.~S.
  {Vaucher}, and A.~H.~M. {van Roermund}, ``A 60 {GHz} phase shifter integrated
  with {LNA} and {PA} in 65 nm {CMOS} for phased array systems,'' \emph{IEEE
  Journal of Solid-State Circuits}, vol.~45, no.~9, pp. 1697--1709, Sep. 2010.

\bibitem{Alkhateeb}
A.~{Alkhateeb}, Y.~{Nam}, J.~{Zhang}, and R.~W. {Heath}, ``Massive {MIMO}
  combining with switches,'' \emph{IEEE Wireless Commun. Lett.}, vol.~5, no.~3,
  pp. 232--235, Jun. 2016.

\bibitem{Buzzi}
S.~{Buzzi} and C.~{D'Andrea}, ``Are mmwave low-complexity beamforming
  structures energy-efficient? analysis of the downlink {MU-MIMO},'' in
  \emph{IEEE Globecom Workshops (GC Wkshps)}, Dec. 2016, pp. 1--6.

\bibitem{Sah}
A.~K. {Sah} and A.~K. {Chaturvedi}, ``Quasi-orthogonal combining for reducing
  {RF} chains in massive {MIMO} systems,'' \emph{IEEE Commun. Lett.}, vol.~6,
  no.~1, pp. 126--129, Feb. 2017.

\bibitem{Payami}
S.~{Payami}, N.~{Mysore Balasubramanya}, C.~{Masouros}, and M.~{Sellathurai},
  ``Phase shifters versus switches: An energy efficiency perspective on hybrid
  beamforming,'' \emph{IEEE Wireless Commun. Lett.}, vol.~8, no.~1, pp. 13--16,
  Feb. 2019.

\bibitem{Gharavi}
M.~{Gharavi-Alkhansari} and A.~B. {Gershman}, ``Fast antenna subset selection
  in {MIMO} systems,'' \emph{IEEE Trans. Signal Process.}, vol.~52, no.~2, pp.
  339--347, Feb. 2004.

\bibitem{Choi}
J.~{Choi}, B.~L. {Evans}, and A.~{Gatherer}, ``Resolution-adaptive hybrid
  {MIMO} architectures for millimeter wave communications,'' \emph{IEEE Trans.
  on Signal Process.}, vol.~65, no.~23, pp. 6201--6216, Dec. 2017.

\bibitem{Wang}
Z.~{Wang}, M.~{Li}, X.~{Tian}, and Q.~{Liu}, ``Iterative hybrid precoder and
  combiner design for mmwave multiuser {MIMO} systems,'' \emph{IEEE Comm.
  Lett.}, 2017.

\bibitem{Liang}
L.~{Liang}, W.~{Xu}, and X.~{Dong}, ``Low-complexity hybrid precoding in
  massive multiuser {MIMO} systems,'' \emph{IEEE Commun. Lett.}, vol.~3, no.~6,
  pp. 653--656, Dec. 2014.

\bibitem{Xiao}
J.~{Li}, L.~{Xiao}, X.~{Xu}, and S.~{Zhou}, ``Robust and low complexity hybrid
  beamforming for uplink multiuser mmwave {MIMO} systems,'' \emph{IEEE Commun.
  Lett.}, 2016.

\bibitem{Golub}
G.~H. Golub and C.~F.~V. Loan, \emph{Matrix computations}.\hskip 1em plus 0.5em
  minus 0.4em\relax Baltimore, MD, USA: Johns Hopkins University Press, 2012.

\bibitem{Bengtsson}
E.~{Björnson}, M.~{Bengtsson}, and B.~{Ottersten}, ``Optimal multiuser
  transmit beamforming: A difficult problem with a simple solution structure
  [{L}ecture {N}otes],'' \emph{IEEE Signal Process. Mag.}, 2014.

\bibitem{Huang}
E.~{Zhang} and C.~{Huang}, ``On achieving optimal rate of digital precoder by
  {RF}-baseband codesign for {MIMO} systems,'' in \emph{IEEE 80th Veh. Technol.
  Conf.}, Sep. 2014.

\bibitem{Kung}
X.~{Zhang}, A.~F. {Molisch}, and K.~{Sun-Yuan }, ``Variable-phase-shifter-based
  {RF}-baseband codesign for {MIMO} antenna selection,'' \emph{IEEE Trans.
  Signal Process.}, vol.~53, no.~11, pp. 4091--4103, 2005.

\end{thebibliography}
	\end{document}